\documentclass[reprint, aps, prd, amsfonts, amssymb, amsmath, showkeys, nofootinbib, superscriptaddress, twoside]{revtex4-2}
\usepackage[english]{babel}
\usepackage[utf8]{inputenc}
\usepackage[colorinlistoftodos, color=green!40, prependcaption]{todonotes}
\usepackage{amsthm}
\usepackage{mathtools,amssymb,bm,bbm}
\usepackage{tensor}
\usepackage{microtype}
\usepackage{braket}
\usepackage{dcolumn}
\usepackage{xcolor}
\usepackage{graphicx}
\usepackage{adjustbox}
\usepackage{placeins}
\usepackage[T1]{fontenc}
\usepackage{lipsum}
\usepackage{csquotes}
\usepackage{slashed}
\usepackage{xspace}
\usepackage[pdftex, pdftitle={Article}, pdfauthor={Author}]{hyperref} % For hyperlinks in the PDF
\usepackage[title]{appendix}
\usepackage{orcidlink}
\usepackage{graphicx}
\usepackage{dcolumn}
\usepackage{subcaption}
\usepackage{mwe}

\newtheoremstyle{mystyle}%                % Name
  {}%                                     % Space above
  {}%                                     % Space below
  {\itshape}%                                     % Body font
  {}%                                     % Indent amount
  {\bfseries}%                            % Theorem head font
  {.}%                                    % Punctuation after theorem head
  { }%                                    % Space after theorem head, ' ', or \newline
  {\thmname{#1}\thmnumber{ #2}\thmnote{ (#3)}}%                                     % Theorem head spec (can be left empty, meaning `normal')

\theoremstyle{mystyle}

\DeclarePairedDelimiter\abs{\lvert}{\rvert}%
\DeclarePairedDelimiter\norm{\lVert}{\rVert}%
\makeatletter
\let\oldabs\abs
\def\abs{\@ifstar{\oldabs}{\oldabs*}}
\let\oldnorm\norm
\def\norm{\@ifstar{\oldnorm}{\oldnorm*}}
\makeatother

\renewcommand\bra[1]{\langle{#1}|}
\makeatletter
\renewcommand\ket[1]{%
  \@ifnextchar\bra{\k@t{#1}\!}{\k@t{#1}}%
}
\newcommand\k@t[1]{{|{#1}\rangle}}
\makeatother

\newcommand{\qb}{\textit{q}\textsc{Bounce}\xspace}

\begin{document}

\title{Decoherence-Free Entropic Gravity for Dirac Fermion}

\author{Eric J. Sung\orcidlink{0000-0003-4437-1068}}
\affiliation{Tulane University, New Orleans, LA 70118, USA}
\email{jsung2@tulane.edu}

\author{Andre G. Campos\orcidlink{0000-0003-2923-4647}}
\affiliation{Max Planck Institute for Nuclear Physics, Heidelberg 69117, Germany}
\email{agontijo@mpi-hd.mpg.de}

\author{Hartmut Abele\orcidlink{0000-0002-6832-9051}}
\affiliation{Technische Universitat Wien, Atominstitut, Stadionallee 2, 1020 Wien, Austria}
\email{hartmut.abele@tuwien.ac.at}

\author{Denys I. Bondar\orcidlink{0000-0002-3626-4804}}
\email{dbondar@tulane.edu}
\affiliation{Tulane University, New Orleans, LA 70118, USA}

\date{\today}

\begin{abstract}
The theory of entropic gravity conjectures that gravity emerges thermodynamically rather than being a fundamental force. One of the main criticisms of entropic gravity is that it would lead to quantum massive particles losing coherence in free fall, which is not observed experimentally. This criticism was refuted in [Phys. Rev. Res. {\bf 3}, 033065 (2021)], where a nonrelativistic master equation modeling gravity as an open quantum system interaction demonstrated that in the strong coupling limit, coherence could be maintained and reproduce conventional free-fall dynamics. Moreover, the nonrelativistic master equation was shown to be fully compatible with the \qb experiment for ultracold neutrons. Motivated by this, we extend these results to gravitationally accelerating Dirac fermions. We achieve this by using the Dirac equation in Rindler space and modeling entropic gravity as a thermal bath thus adopting the open quantum systems approach as well. We demonstrate that in the strong coupling limit, our entropic gravity model maintains quantum coherence for Dirac fermions. In addition, we demonstrate that spin is not affected by entropic gravity. We use the Foldy-Wouthysen transformation to demonstrate that it reduces to the nonrelativistic master equation, supporting the entropic gravity hypothesis for Dirac fermions. Also, we demonstrate how antigravity seemingly arises from the Dirac equation for free-falling antiparticles but use numerical simulations to show that this phenomenon originates from zitterbewegung thus not violating the equivalence principle. 
\end{abstract}

% Also, we demonstrate how anti-gravity seemingly arises from the Dirac equation for free-falling antiparticles but we argue, using numerical simulations, that this phenomenon originates from zitterbewegung and show that the equivalence principle is not violated. 

%\keywords{first keyword, second keyword, third keyword}

\renewcommand\theequation{\arabic{section}.\arabic{equation}}
\counterwithin*{equation}{section}

\maketitle

\section{Introduction}

One of the greatest challenges in modern physics is arguably the unification of gravity and quantum mechanics. Due to the enormous theoretical and experimental success of quantizing three of the four fundamental forces, it is widely assumed that gravity can be quantized as well. However, current hypothetical theories of quantum gravity are plagued with a multitude of problems. This motivates the development of alternative theories of gravity, with entropic gravity being one of them. 

Verlinde's theory of entropic gravity \cite{verlinde_origin_2011} proposes that gravity is an entropic force that arises as a consequence of a system moving toward the direction of maximal entropy, essentially making gravity a thermodynamically emergent, rather than a fundamental, force. If true, this theory would topple the long-standing cherished assumption that gravity has a quantum origin. However, this theory has been criticized for various reasons \cite{kobakhidze_gravity_2011,kobakhidze_once_2011,gao_is_2011,visser_conservative_2011}, with one of the most prominent criticisms being that entropic gravity would couple too strongly and thus destroy quantum coherence \cite{visser_conservative_2011}. This argument, however, was refuted in \cite{schimmoller_decoherence-free_2020}, where a nonrelativistic \emph{decoherence-free entropic gravity} (DFEG) Lindblad master equation was proposed that modeled entropic gravity as an external reservoir coupled to a massive particle with a free dimensionless coupling constant $\sigma$ \cite[Eq.~(5)]{schimmoller_decoherence-free_2020}. The DFEG model predicts that in the strong coupling limit $\sigma \rightarrow \infty$, quantum coherence was still maintained while also recovering Newtonian gravity. This was further supported with an entropic gravity interpretation of the \qb experiment \cite[Eq.~(18)]{schimmoller_decoherence-free_2020} and demonstrating that the DFEG model reproduced the results of the \qb experiment \cite{cronenberg_acoustic_2018} for ultracold neutrons as long as the coupling constant $\sigma \gtrsim 250$. 

In this paper, emboldened by the success of the nonrelativistic theory, we extend the DFEG to Dirac fermions. Our motivation is based on the simple fact that neutrons, a primary subject for experimental gravitational studies, are spin half fermions which are best described by the Dirac equation. Simultaneous description of gravity and Dirac fermions is currently best captured in the \emph{ad hoc} formalism of quantum physics in curved spacetime; thus, a Dirac DFEG model employing this formalism would provide a deeper insight into entropic gravity. We find that spin is not changed in our Dirac DFEG model; thus our model does not conflict with the weak equivalence principle.

%In this paper, emboldened by the success of the nonrelativistic theory,  we extend the DFEG to Dirac fermions. Our motivation is based on the simple fact that neutrons, a primary subject for experimental gravitational studies, are spinors that are best described by the Dirac equation. Simultaneous description of gravity and Dirac fermions is currently best captured in the ad-hoc formalism of quantum physics in curved spacetime; thus, a Dirac DFEG model employing this formalism would provide a deeper insight into the nature of entropic gravity for Dirac fermions. In addition, an entropic gravity model utilizing the Dirac equation allows for insight into the role of quantum spin in an entropic theory of gravity so another aim of this paper is to explore this relationship. 

As explained in \cite{visser_conservative_2011} and demonstrated in \cite{schimmoller_decoherence-free_2020}, the theory of entropic gravity allows for gravity to be modeled as an external thermal reservoir and its interaction with massive particles can be modeled as an open quantum system. To this end, we model entropic gravity by utilizing the theory of open quantum systems via the Lindblad master equation approach. The sheer versatility and success of the Lindblad master equation in nonrelativistic open quantum systems is exemplified by the breadth of applications such as in quantum information  \cite{decoherence_free_subspaces_lidar_chuang,entangled_states_quantum_markov_process_kraus,quantum_computation_and_information_nielsen_chuang_book}, condensed matter physics \cite{open_xxz_spin_chain_prosen_toma,spin_models_of_dissipative_quantum_glasses_olmos,manzano_quantum_2016}, quantum to classical transition \cite{habib_emergence_2006,decoherence_zurek,Adler_2007,csl_model_ghirardi_1990,carlesso_present_2022,bassi_gravitational_2017}, and even in the study of quark-gluon plasmas \cite{quarkonium_suppression_Brambilla,quark_master_equation_lindblad_form_akamatsu}. The theory of open quantum systems also provides a natural framework for studying quantum decoherence, particularly gravitational decoherence \cite{sanchez_gomez1995GravitationalFD,power_decoherence_2000,breuer_metric_2009,blencowe_effective_2013,anastopoulos_master_2013,pikovski_universal_2015} (see Ref.~\cite{bassi_gravitational_2017} for a thorough review); thus this framework is ideal for our work in studying decoherence in Dirac fermions. 

The obtained Dirac DFEG model is physically validated by the fact that in the nonrelativistic limit, it reduces to the aforementioned nonrelativistic DFEG model~\cite{schimmoller_decoherence-free_2020}. Since the latter is compatible with the \qb experiment, so is the Dirac model.

The rest of the paper is organized as follows: In Sec.~\ref{section: quantum physics in accelerated frames}, for completeness and self-consistency, we rederive the geometry of physics in accelerated frames, which we use to derive the Dirac equation in Rindler space. In Sec.~\ref{section: antigravity and ehrenfest theorems}, we derive the Ehrenfest theorems for the Dirac equation in the Rindler space and discuss anti-gravity which automatically follows for antiparticles. We use numerical simulations to show that this anti-gravity phenomenon originates from \emph{zitterbewegung} and that the equivalence principle is not violated. Then in Sec.~\ref{section: relativistic dfeg master equation}, we use the Dirac equation in Rindler space \eqref{low energy rind ham} from Sec.~\ref{section: quantum physics in accelerated frames} and the Ehrenfest theorems from Sec.~\ref{section: antigravity and ehrenfest theorems} to formulate the DFEG master equation for Dirac fermions~\eqref{relativistic dfeg master eq}, which is the main result of this work. We demonstrate that by increasing the coupling constant $\sigma$, the master equation \eqref{relativistic dfeg master eq} can achieve arbitrarily low decoherence and reduces to the Dirac equation in a linear gravitational potential in the $\sigma \rightarrow \infty$ limit. In addition, we show that the spin is preserved by entropic gravity. In Sec.~\ref{section: boundary condition of bouncing dirac fermion}, we choose and rederive the boundary conditions from Ref.~\cite{boulanger_bound_2006} which will be used to model the \qb experiment and give some insight into the difficulty of formulating boundary conditions for the Dirac equation. In Sec.~\ref{section: relativistic qbounce hamiltonian}, we relativistically model the \qb experiment using the Ehrenfest theorems of the Dirac equation in Rindler space and the adopted boundary condition. We then use the results of Sec.~\ref{section: relativistic dfeg master equation} to construct the relativistic DFEG master equation for the \qb experiment. In Sec.~\ref{section: nonrelativistic limit}, we demonstrate that in the nonrelativistic limit, our relativistic results correctly reduce to their nonrelativistic counterparts in \cite{schimmoller_decoherence-free_2020}. In Appendix \ref{appendix: decoherence property}, we prove that our entropic gravity model is decoherence-free. In Appendix \ref{appendix: rindler dirac equation: eigenfunctions and eigenenergies}, we solve the Dirac equation in Rindler space to find its spin-dependent energy levels and eigenspinors. Then in Appendix \ref{appendix: normalization rindler wavefunction}, we calculate the normalization constant. We also provide a brief discussion of the nature of spin-gravity coupling and recent experiments on it.

Throughout this paper, we adopt the usual Einstein summation convention with Greek indices running from the temporal and spatial indices $0$--$3$ and Latin indices running only the spatial indices $1$--$3$, unless stated otherwise. The binary operations $[\cdot,\cdot]$ and $\{\cdot,\cdot \}$ denote the commutator and anticommutator, respectively. We use the ``mostly negative'' metric signature $(+,-,-,-)$ and denote the Minkowski and curved metrics as $\eta_{\mu\nu}$ and $g_{\mu\nu} \equiv g_{\mu\nu}(x)$, respectively. We let $\mathbbm{1}_{n}$ and $\sigma^{i}$ denote the $n\times n$ identity and $2\times 2$ Pauli matrices, respectively. Unless stated otherwise, we use the $4\times 4$ gamma matrices $\gamma^{\mu}$ in the Dirac representation 
\begin{align}
    \gamma^{0} = \begin{pmatrix}
    \mathbbm{1}_{2} & 0 \\
    0 & -\mathbbm{1}_{2}
    \end{pmatrix}, \quad \gamma^{i} = \begin{pmatrix}
    0 & \sigma^{i} \\
    -\sigma^{i} & 0
    \end{pmatrix}, 
\end{align}
which obeys the Clifford algebra in Minkowski space 
\begin{align}
    \{\gamma^{\mu}, \gamma^{\nu} \} &= 2\eta^{\mu\nu}. \label{clifford algebra dirac}
\end{align}
Then we have $\gamma_{5} = i\gamma^{0}\gamma^{1}\gamma^{2}\gamma^{3}$, $\alpha_{i} = \gamma^{0} \gamma^{i}$ and $\beta = \gamma^{0}$. We choose the $z$-direction for our linear equations.

\section{Quantum Physics In Accelerated Frames} \label{section: quantum physics in accelerated frames}

We begin with a re-derivation of the geometry of physics in an accelerated frame that will be used to derive the spin connection in an accelerated frame. Then we proceed to derive the Dirac equation in an accelerated frame. We shall show that by various coordinate transformations, the metric and coordinates we derive are equivalent to previous formulations. The results developed in this section provide the necessary background for formulating the entropic gravity model for Dirac fermions. 

\subsection{Rindler Space}

The \qb experiment \cite{cronenberg_acoustic_2018} measured the effect of Earth's gravity on ultracold neutrons by using gravity resonance spectroscopy to induce transitions between the quantum states of the bouncing ball via a vibrating mirror. In the nonrelativistic regime, this is physically modeled as a neutron bouncing in the $z$-direction on a fixed surface due to the influence of a linear gravitational potential $mg\hat{z}$, where $g$ is the gravitational acceleration near Earth's surface. To get the particle to ``bounce,'' one imposes the Dirichlet boundary condition and finds that the energy levels of the bouncing particle are proportional to the Airy function zeros \cite{sakurai}. 

In our relativistic interpretation of the \qb experiment, we imagine a relativistic massive Dirac fermion moving with uniform acceleration in the $z$-direction under the influence of the Earth's gravity, hitting a vibrating mirror, and achieving a similar ``bouncing ball'' state \cite{rohim_relativistic_2021,greiner_quantum_1985}. This means that we are working with accelerated frames, and thus we cannot simply use the usual Dirac equation in Minkowski space since this equation is only valid for inertial frames. Hence, following Refs.~\cite{misner1973gravitation,hehl_inertial_1990}, we return to the geometric foundations and rederive the appropriate metric tensor $g_{\mu\nu}$ to describe physics in accelerated frames. 

Suppose that in an inertial frame with Minkowski coordinates $x^{\mu^{\prime}}=(x^{0^{\prime}}, x^{1^{\prime}}, x^{2^{\prime}}, x^{3^{\prime}})$, an observer moves with an arbitrary, finite proper three-acceleration $\textbf{a}(\tau)$ parametrized by their proper time $\tau$. Additionally, let $u^{\mu^{\prime}}$ be the four-velocity of the observer relative to the inertial frame. In this inertial frame, the accelerated observer carries a tetrad frame $e_{\alpha}(\tau)$ such that
\begin{align}
    e_{0} &= c^{-1} u^{\mu^{\prime}}, \label{rest frame condition} \\
    e_{\mu} \cdot e_{\nu} &= \eta_{\mu \nu}, \label{orthonormal condition}
\end{align}
namely, the observer's basis vectors form a rest frame at each instant, and the tetrads are orthonormal, respectively. We also demand that the tetrads be nonrotating in the sense that only the timelike plane of the four-velocity and four-acceleration is rotated while all other planes are excluded from rotation \cite{misner1973gravitation}. Then the orthonormal tetrad frame $e_{\alpha}$ is Fermi-Walker transported according to 
\begin{equation}
    \frac{d e_{\alpha}}{d \tau} = \bm{\Omega} \cdot e_{\alpha}, \label{trans law}
\end{equation}
where 
\begin{align}
    \Omega^{\mu\nu} &= \frac{(a^{\mu}u^{\nu} - a^{\nu}u^{\mu})}{c^{2}},\label{rot tensor}
\end{align}
is the antisymmetric rotation tensor with $a^{\mu}$ being the observer's four-acceleration. Now let $z^{\mu^{\prime}}(\tau)$ be the displacement vector from the inertial frame to the observer's position $\mathcal{P}(\tau)$. At each point $\mathcal{P}(\tau)$ on the observer's worldline, let the observer have the spacelike basis vectors $e_{i^{\prime}}$, and then these spacelike basis vectors define a spacelike hyperplane with the spatial components of the tetrad being $e_{i^{\prime}}=e_{i}$ \cite{misner1973gravitation}. We then use the spatial tetrads $e_{i}$ to construct the observer's ``local coordinates'' $x^{\mu}=(x^{0},x^{1},x^{2},x^{3})$ at the origin where $x^{i} \equiv \mathbf{x}$ are the Cartesian coordinates in the hyperplane and $x^{0}\equiv ct \equiv c\tau$ \cite{hehl_inertial_1990,misner1973gravitation}. Then each event on the hyperplane has coordinates
\begin{align}
    x^{\mu^{\prime}}(\tau) &= x^{i} (e_{i}(\tau))^{\mu^{\prime}} + z^{\mu^{\prime}}(\tau). \label{hyperplane formula}
\end{align}

Suppose now the observer moves in the $x^{3}$-direction with uniform acceleration $\mathbf{a}=(0,0,g)$ and $x^{1^{\prime}}=x^{2^{\prime}}=0$ in the inertial frame. Then the observer's four-velocity and four-acceleration, relative to the inertial frame, satisfy 
\begin{align}
    u^{\mu^{\prime}}u_{\mu^{\prime}}=c^{2}, \quad a^{\mu^{\prime}}a_{\mu^{\prime}}=-g^{2}, \quad u^{\mu^{\prime}}a_{\mu^{\prime}} = 0. \label{four velocity and acceleration conditions}
\end{align}
The third equation in Eqs.~\eqref{four velocity and acceleration conditions} implies that $a^{0^{\prime}}=0$ in the observer's rest frame, i.e., $e_{0}=c^{-1} u^{\mu^{\prime}}$ at that instant. Solving Eqs.~\eqref{four velocity and acceleration conditions} for $x^{0^{\prime}}$ and $x^{3^{\prime}}$ yields
\begin{align}
    x^{0^{\prime}} = \frac{c^{2}}{g}\sinh{(g\tau/c)}, \quad x^{3^{\prime}} = \frac{c^{2}}{g}\cosh{(g\tau/c)},
\end{align}
then the displacement vector is
\begin{align}
    z^{\mu^{\prime}}(\tau) &= \left(\frac{c^{2}}{g}\sinh{(g\tau/c)},0,0,\frac{c^{2}}{g}\cosh{(g\tau/c)}\right). \label{observers worldline}
\end{align}
To find the tetrad basis carried by the observer, we note that since $e_{1}$ and $e_{2}$ are invariant under Lorentz transformations in the $x^{3}$-direction, $e_{1}=e_{1^{\prime}}$ and $e_{2}=e_{2^{\prime}}$ must be the unit basis vectors. Since $e_{0}=c^{-1} u^{\mu^{\prime}}$, we use the orthonormality \eqref{orthonormal condition} and nonrotating conditions to find that $e_{3}=g^{-1} a^{\mu^{\prime}}$, namely, $e_{3}$ is parallel to the acceleration. Thus the tetrad basis carried by the observer is \cite{misner1973gravitation}
\begin{align}
    (e_{0})^{\mu^{\prime}} &= (\cosh{(g\tau /c)},0,0,\sinh{(g\tau /c)}), \notag \\
    (e_{1})^{\mu^{\prime}} &= (0,1,0,0), \notag \\
    (e_{2})^{\mu^{\prime}} &= (0,0,1,0), \notag \\
    (e_{3})^{\mu^{\prime}} &= (\sinh{(g\tau /c)},0,0,\cosh{(g\tau /c)}). \label{basis tetrads}
\end{align}
It can be shown that tetrads \eqref{basis tetrads} are nonrotating and obey conditions \eqref{rest frame condition}-\eqref{orthonormal condition}. By using Eq.~\eqref{hyperplane formula} with vector \eqref{observers worldline} and tetrads \eqref{basis tetrads}, we get the components of $x^{\mu^{\prime}}$ 
\begin{align}
    x^{0^{\prime}} &= \left(x^{3} + \frac{c^{2}}{g} \right) \text{sinh}(g\tau/c), \notag \\
    x^{1^{\prime}} &= x^{1}, \notag \\
    x^{2^{\prime}} &= x^{2}, \notag \\
    x^{3^{\prime}} &= \left(x^{3} + \frac{c^{2}}{g} \right)\text{cosh}(g\tau/c), \label{kottler moller coordinates}
\end{align}
with the Minkowski line element 
\begin{align}
    ds^{2} &= \eta_{\mu^{\prime} \nu^{\prime}} dx^{\mu^{\prime}} dx^{\nu^{\prime}} \notag \\
    &= \left(1 + \frac{gx^{3}}{c^{2}} \right)^{2} (dx^{0})^{2} -(dx^{1})^{2}- (dx^{2})^{2} - (dx^{3})^{2}. \label{rindler line element}
\end{align}
If we now define the new timelike and spacelike comoving coordinates 
\begin{align}
    v = \frac{g \tau}{c}, \quad u = x^{3} + \frac{c^{2}}{g}, \label{rindler coordinates}
\end{align}
respectively, we get 
\begin{align}
    x^{0^{\prime}} &= u\, \text{sinh}(v), \notag \\
    x^{1^{\prime}} &= x^{1}, \notag \\
    x^{2^{\prime}} &= x^{2}, \notag \\
    x^{3^{\prime}} &= u\, \text{cosh}(v), \label{actual rindler coordinates}
\end{align}
where $v=\text{artanh}(x^{0^{\prime}}/x^{3^{\prime}})$ with $u \in [0,\infty)$ and $v \in (-\infty,\infty)$. These new comoving coordinates $x^{\mu}=(v,x^{1},x^{2},u)$ are the famous Rindler coordinates and due to the bounds on $u$ and $v$, we are specifically working with the right Rindler wedge in Minkowski space \cite{ueda_entanglement_2021, greiner_quantum_1985}. The trajectory of the uniformly accelerated observer is then
\begin{equation}
    (x^{3^{\prime}})^{2} - (x^{0^{\prime}})^{2} = u^{2}= \left(x^{3} + \frac{c^{2}}{g} \right)^{2},
\end{equation}
thus the observer's worldline is a hyperbola in Minkowski space \cite{greiner_quantum_1985,misner1973gravitation}. The Minkowski line element in the Rindler coordinates is
\begin{align}
ds^{2} &=g_{\mu \nu} dx^{\mu} dx^{\nu} \notag \\
&= u^{2}dv^{2} -(dx^{1})^{2}- (dx^{2})^{2} - du^{2}, \label{rindler line elem} 
\end{align}
which gives the Rindler space metric tensor
\begin{align}
    g_{\mu\nu} &= \text{diag}(u^{2},-1,-1,-1). \label{rindler metric}
\end{align}
To aid our work in the next subsection, we find the tetrads $e_{\alpha}$ in terms of the Rindler coordinates \eqref{actual rindler coordinates}. Using the Rindler metric \eqref{rindler metric} and the orthonormality relation
\begin{align}
    e_{\mu} \cdot e_{\nu} &= g_{\mu\nu},
\end{align}
we find that
\begin{align}
    (e_{0})^{\mu} = u^{-1} \delta^{\mu}_{0}, \quad (e_{i})^{\mu}= \delta^{\mu}_{i}, \label{rindler tetrad}
\end{align}
where $\delta^{\mu}_{\nu}$ is the Kronecker delta function.

It should be noted that coordinates \eqref{actual rindler coordinates} are the original Rindler coordinates \cite{rindler_1966} while the coordinates \eqref{kottler moller coordinates} that we used to derive the actual Rindler coordinates are called the Kottler-Møller coordinates \cite{kottler_1916,moller_1943,misner1973gravitation}. There exist many other equivalent coordinate systems for describing uniform acceleration in Minkowski space that, of course, also lead to hyperbolic trajectories. Another popular choice of coordinates describing the dynamics of a uniformly accelerated observer can be shown by a coordinate transformation on the Rindler position variable $u$ to
\begin{align}
    u &= \frac{c^{2}}{g} e^{g\xi/c^{2}}, \label{spatial radar lass coordinates}
\end{align}
where $\xi$ is a spatial variable, which turns coordinates \eqref{actual rindler coordinates} into
\begin{align}
    x^{0^{\prime}} &= \frac{c^{2}}{g} e^{g\xi/c^{2}}\, \text{sinh}(g\tau/c), \notag \\
    x^{1^{\prime}} &= x^{1}, \notag \\
    x^{2^{\prime}} &= x^{2}, \notag \\
    x^{3^{\prime}} &= \frac{c^{2}}{g} e^{g\xi/c^{2}}\, \text{cosh}(g\tau/c), \label{radar lass coordinates}
\end{align}
where we have opted to use the explicit form of the Rindler temporal variable $v$. This choice \eqref{radar lass coordinates} is called the Radar or Lass coordinates \cite{lass_1963}, and it gives the Radar or Lass line element and metric 
\begin{align}
    ds^{2} &= e^{2g\xi/c^{2}} (dx^{0})^{2}  -(dx^{1})^{2}- (dx^{2})^{2}- e^{2g\xi/c^{2}}d\xi^{2}, \\
    g_{\mu\nu} &= \text{diag}(e^{2g\xi/c^{2}},-1,-1,-e^{2g\xi/c^{2}}),
\end{align}
respectively, (and ultimately the Dirac equation) used in other literature (see Refs.~\cite{rohim_relativistic_2021,crispino_unruh_2008,ueda_entanglement_2021}). Conversely, one could start with the spatial Radar coordinate \eqref{spatial radar lass coordinates} and in the weak gravitational limit, namely, $g\xi/c^{2} \ll 1$, expand the coordinate up to first order $e^{g\xi/c^{2}} = 1+g\xi/c^{2}+\ldots$ to get the Kottler-Møller coordinates \eqref{kottler moller coordinates} and ultimately the Rindler coordinates \eqref{actual rindler coordinates}. At the end of the following subsection, we explain our rationale for choosing coordinates \eqref{rindler coordinates} as the preferred Rindler coordinates.

\subsection{Dirac Equation in Rindler Space}

With the geometric preliminaries firmly established, we now turn our attention to the Dirac equation. Recall that the (inertial) Dirac equation in Minkowski space is
\begin{align}
    (i\hbar\gamma^{\mu}\partial_{\mu} - mc) \Psi = 0. \label{inertial dirac eq}
\end{align}
To incorporate the geometric information encoded in the Rindler space metric tensor \eqref{rindler metric}, we use the minimal coupling and Einstein equivalence principles \cite{di_casola_nonequivalence_2015} on Eq.~\eqref{inertial dirac eq} to get the Dirac equation in curved spacetime \cite{carroll2003spacetime}
\begin{equation}
    (i\hbar\gamma^{\mu}_{R}\nabla_{\mu} - mc) \Psi = 0,
\end{equation}
with the covariant derivative 
\begin{align}
    \nabla_{\mu} &= \partial_{\mu} + \Gamma_{\mu}, \\
    \partial_{\mu} &\equiv (\partial_{0},\partial_{1},\partial_{2},\partial_{3}) \equiv (\partial_{v},\partial_{1},\partial_{2},\partial_{u}),
\end{align}
and spin connection \cite{mashhoon_quantum_2006,obukhov_gravitational_2001}
\begin{equation}
    \Gamma_{\mu} = \frac{1}{4}\gamma_{R \, \nu} \left(\frac{\partial \gamma^{\nu}_{R}}{\partial x^{\mu}} + \Gamma\indices{^{\nu}_{\lambda \mu}} \gamma^{\lambda}_{R} \right) = \frac{1}{4}\gamma_{R \, \nu}D_{\mu}\gamma^{\nu}_{R},
\end{equation}
where 
\begin{align}
    \gamma^{\mu}_{R} \equiv \gamma^{\mu}_{R}(x) = (e_{\nu})^{\mu}\gamma^{\nu}, \label{rindler gamma}
\end{align}
are the ``curved'' gamma matrices which obey the curved Clifford algebra
\begin{equation}
    \{ \gamma_{R}^{\mu}(x), \gamma_{R}^{\nu}(x) \} = 2g^{\mu \nu}(x). \label{clifford alg rindler}
\end{equation}
To express the curved gamma matrices $\gamma^{\mu}_{R}$ in terms of the ``flat'' gamma matrices $\gamma^{\mu}$, we use Eq.~\eqref{rindler gamma} with the Rindler tetrads \eqref{rindler tetrad} and the curved Clifford algebra \eqref{clifford alg rindler} to get
\begin{alignat}{3}
    \gamma^{0}_{R} &= \frac{1}{u} \gamma^{0} , \quad &&\gamma_{R \,0} = u \gamma_{0}, \\
    \gamma^{i}_{R} &= \gamma^{i}, \quad &&\gamma_{R \, i} = \gamma_{i}.
\end{alignat}
Then the spin connection in Rindler space is
\begin{equation}
    \Gamma_{\mu} = \left(\frac{1}{2}\gamma^{0}\gamma^{3},0,0,0 \right),
\end{equation}
and the Dirac equation in Rindler space is
\begin{align}
    &\left[i\hbar\gamma^{0} \partial_{v} + i\hbar u \gamma^{i}\partial_{i} + \frac{i\hbar}{2}\gamma^{3} - mcu \right] \Psi=0. \label{3d cov rind dir eq}
\end{align}
Multiplying by $\gamma^{0}$ on the left of Eq.~\eqref{3d cov rind dir eq} and rearranging terms yields the full Rindler space Dirac equation
\begin{align}
    i\hbar\partial_{v} \Psi &=  \left[-i\hbar u \alpha_{i} \partial_{i} - \frac{i\hbar}{2} \alpha_{3} + \beta mcu \right] \Psi \equiv \hat{H}_{\mathbf{R}} \Psi. \label{3d dirac ham}
\end{align}
From the Rindler coordinates \eqref{rindler coordinates}, we deduce that the Rindler position and momentum operators are 
\begin{alignat}{3}
    &\hat{\mathbf{x}} = (\hat{x}_{1},\hat{x}_{2},\hat{u}) &&\rightarrow \qquad \mathbf{x} &&= (x_{1},x_{2},u), \notag \\
    &\hat{\mathbf{p}} = (\hat{p}_{1},\hat{p}_{2},\hat{p}_{u}) &&\rightarrow -i\hbar\partial_{i} &&= (-i\hbar\partial_{1},-i\hbar\partial_{2}, -i\hbar \partial_{u}), \label{rind position momentum ops}
\end{alignat}
respectively, which obey the canonical commutation relations
\begin{align}
    [\hat{x}_{a},\hat{p}_{b}]=i\hbar \delta_{a,b}, \quad [\hat{u},\hat{p}_{u}]=i\hbar, \quad a,b=1,2,
\end{align}
so the full Rindler Hamiltonian in operator form is
\begin{align}
    \hat{H}_{\mathbf{R}} &=  \alpha_{a}\hat{u}  \hat{p}_{a} + \alpha_{3} \hat{u}\hat{p}_{u} - \frac{i\hbar}{2} \alpha_{3} + \beta mc\hat{u}. \label{3d dirac ham op}
\end{align}
Since we are considering linear gravity in the $z$-direction, we drop the other directional terms in Eqs.~\eqref{3d cov rind dir eq} and \eqref{3d dirac ham op} to get the linear Rindler Dirac equation and Hamiltonian
\begin{align}
    i\hbar\partial_{v} \Psi &=  \left[-i\hbar u \alpha_{3} \partial_{u} - \frac{i\hbar}{2} \alpha_{3} + \beta mcu \right] \Psi \equiv \hat{H}_{R} \Psi, \label{rind ham eq} \\
    \hat{H}_{R} &= \alpha_{3} \hat{u} \hat{p}_{u} -\frac{i\hbar}{2}\alpha_{3} + \beta mc\hat{u}, \label{rind ham op form}
\end{align}
respectively.

To express the full Rindler Hamiltonian \eqref{3d dirac ham op} in the observer's coordinates, we first use the Rindler coordinates \eqref{rindler coordinates} to get the inverse operator transformations
\begin{align}
    &\hat{x}_{a}=\hat{x}_{a}, \quad \hat{u} = \hat{x}^{3} + \frac{c^{2}}{g}, \notag \\
    &\hat{p}_{a} = \hat{p}_{a}, \quad \hat{p}_{u} \rightarrow -i\hbar \partial_{u} = -i\hbar \partial_{3} \rightarrow \hat{p}_{3}, \notag \\
    &\hat{E}_{v} \rightarrow i\hbar \partial_{v} = \frac{i\hbar c}{g} \partial_{t} \rightarrow \frac{c}{g} \hat{E}_{t}, \label{inverse coord trans}
\end{align}
then use the inverse transformations \eqref{inverse coord trans} on Hamiltonian \eqref{3d dirac ham op} to get our desired result
\begin{align}
    \hat{H}_{\mathbf{G}} &=c\bm{\alpha} \cdot \hat{\mathbf{p}} + \beta mc^{2}+ \beta m(\mathbf{a} \cdot\hat{\mathbf{x}}) + \frac{1}{2c} \{(\mathbf{a} \cdot \hat{\mathbf{x}}), (\bm{\alpha} \cdot \hat{\mathbf{p}}) \}, \label{3d non FW ham}
\end{align}
where $\mathbf{a}=(0,0,g)$. Note that Hamiltonian \eqref{3d non FW ham} is the nonrotational version of the Hamiltonian in Hehl and Ni \cite[Eq.~(16)]{hehl_inertial_1990}. The linear version of Hamiltonian \eqref{3d non FW ham} is
\begin{align}
    \hat{H}_{G} &=c\alpha_{3} \hat{p}_{3} + \beta mc^{2}+ \beta mg\hat{z} + \frac{g}{2c} \alpha_{3} \{\hat{z}, \hat{p}_{3} \}. \label{non FW ham}
\end{align}
Since our work is concerned with low energy effects, we disregard the negligible fourth redshift term in Eq.~\eqref{non FW ham}, which leaves us with the low energy gravitational Dirac Hamiltonian
\begin{align}
    \hat{H}_{g} &= c\alpha_{3} \hat{p}_{3} + \beta mc^{2} + \beta mg \hat{z}. \label{low energy rind ham}
\end{align}

\begin{figure}
\begin{subfigure}[t]{0.48\textwidth}
\includegraphics{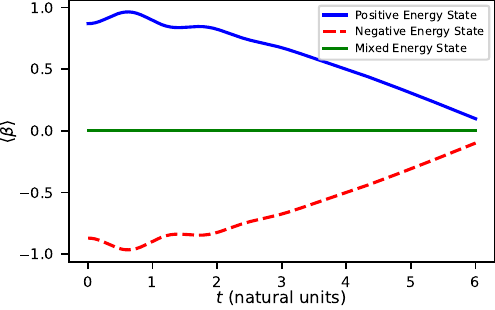}% Here is how to import EPS art
\end{subfigure}
\caption{\label{fig:betaRindler} Time evolution of $\braket{\beta}$ for matter, antimatter, and a mixture of both. We see that $\braket{\beta}$ rapidly goes to zero for both the matter and antimatter states.}
\end{figure}

With the Rindler metric \eqref{rindler metric}, the Rindler space Dirac inner product is
\begin{align}
    \Braket{\Psi_{\Omega,s,\bm{k}_{\perp}} | \Psi_{\Omega^{\prime},s^{\prime},\bm{k}_{\perp}^{\prime}}} &= \int_{\Sigma} d\Sigma_{\mu} \, \overline{\Psi}_{\Omega,s,\bm{k}_{\perp}} \gamma^{\mu}_{R} \Psi_{\Omega^{\prime},s^{\prime},\bm{k}_{\perp}^{\prime}} \notag \\
    &= \delta(\Omega - \Omega^{\prime}) \, \delta_{s,s^{\prime}} \, \delta(\bm{k}_{\perp} - \bm{k}_{\perp}^{\prime})  \label{dirac inner product}
\end{align}
where $d\Sigma_{\mu} = d\Sigma n_{\mu}$ is the spatial volume element on the $v=\text{constant}$ Cauchy hypersurface $\Sigma$, $n_{\mu}$ is the unit vector normal to $\Sigma$, $s=\pm$ is the spin orientation, $\Omega=\omega c/g$ is the dimensionless frequency, $\bm{k}_{\perp}\equiv k^{a}=(k^{1},k^{2},0)$ is the wavevector perpendicular to the direction of acceleration, and $\overline{\Psi}_{\Omega,s,\bm{k}_{\perp}}=\Psi^{\dagger}_{\Omega,s,\bm{k}_{\perp}} \gamma^{0}$ is the adjoint spinor \cite{carroll2003spacetime,crispino_unruh_2008,ueda_entanglement_2021}. 

It is worth mentioning that had we derived Hamiltonian \eqref{3d non FW ham} in the context of a rotating frame with rotation frequency $\bm{\omega}(\tau)$, we would introduce the rotation-angular momentum coupling term $-\bm{\omega} \cdot \hat{\mathbf{J}}$ in Hamiltonian \eqref{3d non FW ham} which represents the coupling of the frame's rotation $\bm{\omega}(\tau)$ to the observer's total angular momentum $\hat{\mathbf{J}} = \hat{\mathbf{L}} + \mathbf{S}$ \cite{hehl_inertial_1990}. The rotation-orbital momentum coupling $\bm{\omega} \cdot \hat{\mathbf{L}}$ creates an effect very reminiscent of the Sagnac effect and induces a phase shift. This Sagnac-like effect has been experimentally verified for neutrons \cite{effect_of_earths_rotation_onphase_shift_neutron_werner}. The rotation-spin angular momentum coupling $\bm{\omega} \cdot \mathbf{S}$ induces a phase shift smaller than the Sagnac-like effect $\bm{\omega} \cdot \hat{\mathbf{L}}$ and was recently observed in neutron interferometry experiments \cite{danner_spin-rotation_2020}.

As mentioned previously, the choice of the Rindler coordinates will lead to slightly different forms of the Rindler Dirac Hamiltonian, and this is most pronounced when the Rindler Hamiltonian is brought to its nonrelativistic limit (see, e.g., Refs.~\cite{soffel_dirac_1980,rohim_relativistic_2021,crispino_unruh_2008,ueda_entanglement_2021}). Our choice of Rindler coordinates \eqref{rindler coordinates} is desirable due to the fact that the observer's local coordinate system \eqref{kottler moller coordinates} is what is actually used in the laboratory \cite{hehl_inertial_1990}. Most importantly, such a choice of Rindler coordinates leads to Hamiltonian \eqref{3d non FW ham} whose terms (along with the rotation-angular momentum coupling terms) have been experimentally verified for neutrons. This gives Hamiltonian \eqref{3d non FW ham}, and the methodology used in its derivation, both theoretical and experimental validity in accurately modeling the behavior of a Dirac fermion in noninertial frames. Since our relativistic interpretation of the \qb experiment uses Dirac fermions, we believe that Hamiltonian \eqref{3d non FW ham}, and its linear version \eqref{non FW ham}, is the physically most appropriate choice.

\section{Zitterbewegung Anti-Gravity and Ehrenfest Theorems} \label{section: antigravity and ehrenfest theorems}

In this section, we derive the Ehrenfest theorems from the low energy Rindler Hamiltonian \eqref{low energy rind ham} which will be used to construct the dissipator that models entropic gravity in Sec.~\ref{section: relativistic dfeg master equation}. We provide numerical simulations of the dynamics of a Dirac fermion with Hamiltonian \eqref{low energy rind ham} and its physical interpretation. In addition, we elaborate on the effect of a \emph{zitterbewegung} induced anti-gravity from our simulations. 

% In this section, we derive the Ehrenfest theorems from the low energy Rindler Hamiltonian \eqref{low energy rind ham} which will be used to construct the dissipator that models entropic gravity in Sec.~\ref{section: relativistic dfeg master equation}. We also give a brief discussion on anti-gravity, which emerges from the Ehrenfest theorems, and how its appearance is due to zitterbewegung. 

% \begin{figure}
% \centering
% \subfloat[]{\includegraphics{fig_x.pdf}\label{fig:fig_x} }\\
% % \caption{Comparing the time evolution of the average position for matter, antimatter, and a mixture of both. Note how the matter and antimatter positions coincide and start to follow classical trajectories. For this simulation plot, we use $x_{\text{init}} = v_{\text{init}} = 0$, $m = 1$, and $g = 0.5$.}
% \subfloat[]{\includegraphics{fig_p.pdf}\label{fig:fig_p}}
% % \caption{Comparing the time evolution of the average momentum for matter, antimatter, and a mixture of both. For this simulation plot, we use $p_{\text{init}} = 0$, $m = 1$, and $g = 0.5$.}
% \caption{Time evolution of the average position \ref{fig:fig_x} and momentum \ref{fig:fig_p} for matter, antimatter, and a mixture of both. We see in Fig.~\ref{fig:fig_x} that the matter and antimatter follow the same trajectory, as prescribed by the equivalence principle.}
% \label{fig:fig_x_p}
% \end{figure}

\begin{figure}
\centering
\begin{subfigure}[t]{0.48\textwidth}
        \includegraphics[width=\textwidth]{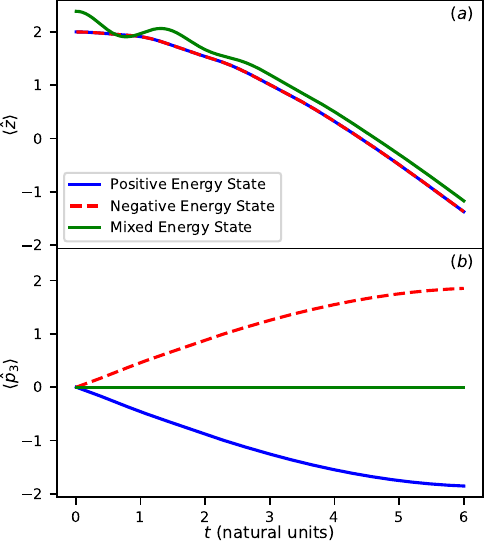}
        \phantomcaption
        \label{fig:fig_x}
    \end{subfigure}
\begin{subfigure}[t]{0\textwidth} % the hidden unwanted image
         \includegraphics[width=\textwidth]{example-image-b}
         \phantomcaption
         \label{fig:fig_p}   
    \end{subfigure}
\caption{Time evolution of the average position (a) and momentum (b) for matter, antimatter, and a mixture of both. We see in Fig.~\ref{fig:fig_x} that the matter and antimatter follow the same trajectory, as prescribed by the equivalence principle.}
\label{fig:fig_x_p}
\end{figure}

To describe the dynamics of a Dirac fermion in a gravitational potential, it is natural to utilize the Ehrenfest theorems of the low energy Rindler Hamiltonian \eqref{low energy rind ham}, which are calculated as \cite{cabrera_operational_2019}
\begin{alignat}{4}
    &\frac{d}{dt} \langle \hat{z}\rangle &=  \phantom{+} \Braket{\frac{\partial \hat{H}_{g}}{\partial \hat{p}_{3}}} &= \phantom{+} c\Braket{\alpha_{3}}, \label{ehren x} \\
    &\frac{d}{dt} \langle \hat{p}_{3} \rangle &= -\Braket{\frac{\partial \hat{H}_{g}}{\partial \hat{z}}} &= - mg\Braket{\beta}. \label{ehren p}
\end{alignat}
Unlike the nonrelativistic Ehrenfest theorems for a linear gravitational potential, Eqs.~\eqref{ehren x}-\eqref{ehren p} depend on the $\alpha_{3}$ and $\beta$ matrices, highlighting the incorporation of antimatter free fall dynamics. To see a Dirac fermion's spin dynamics under Eq.~\eqref{low energy rind ham}, we also calculate its Ehrenfest theorem. Recall that the $4\times 4$ spin observables are
\begin{align}
    \mathbf{S} &= \frac{\hbar}{2} \bm{\Sigma} = \frac{\hbar}{2}\begin{pmatrix}
        \bm{\sigma} & 0 \\
        0 & \bm{\sigma}
    \end{pmatrix}, \label{spin matrix}
\end{align}
which have the commutation relations
\begin{align}
    [\alpha_{i},S_{j}] = i\hbar \varepsilon_{ijk} \alpha_{k}, \quad [\beta,S_{j}]=0, \label{spin comm rel}
\end{align}
where $\varepsilon_{ijk}$ is the Levi-Civita tensor. The first commutation relation in Eqs.~\eqref{spin comm rel} can be deduced using 
\begin{align}
    [\sigma^{i},\sigma^{j}] &= 2i \varepsilon^{ijk} \sigma^{k}.
\end{align}
Then the Ehrenfest theorem is
\begin{align}
    \frac{d}{dt}\braket{S_{3}} &= \frac{1}{i\hbar} \braket{[ S_{3},\hat{H}_{g} ]} = 0,
\end{align}
thus the spin is conserved. We note that the full Rindler Hamiltonian \eqref{non FW ham} conserves spin as well.

\begin{figure}
\centering
\begin{subfigure}[t]{0.48\textwidth}
\includegraphics{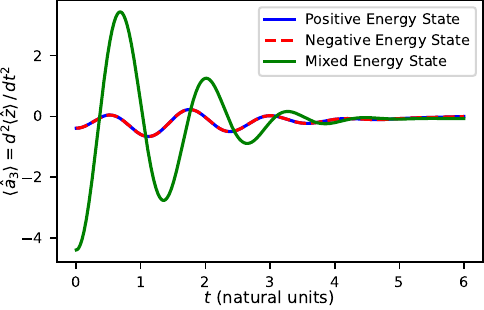}% Here is how to import EPS art
\end{subfigure}
\caption{Time evolution of the average acceleration for matter, antimatter, and a mixture of both. We see that the acceleration for antimatter aligns with matter thus demonstrating that antimatter obeys the equivalence principle.}
\label{fig:fig_a} 
\end{figure}

Using the propagator~\cite{cabrera_dirac_2016}, we numerically solve the Dirac equation~\eqref{low energy rind ham} to understand fermion’s free fall dynamics. We considered  three different initial conditions of the Dirac fermion: positive energy, negative energy, and mixed energy wave packets. To get the positive energy initial state, we first take the Gaussian wave packet (see, e.g., Fig. 2(a) in \cite{cabrera_operational_2019}) centered at zero momentum and position $z_\text{init} = 2$ (natural units $\hbar = c = m =1$ and $g = 0.5$ are employed) and apply the projector \cite[Eq.~(37)]{cabrera_operational_2019} to eliminate negative energy components (i.e., antiparticles). The negative energy initial state, entirely made of antimatter, is obtained similarly by projecting out the positive energy components (i.e., matter). The mixed energy state, centered at the position of $z=2.38$, is made in equal proportion of matter and antimatter. Numerically obtained time evolution of $\braket{\beta}$, $\braket{\hat{z}}$, $\braket{\hat{p}_{3}}$, and $\braket{\hat{a}_{3}} = d^{2} \braket{\hat{z}} / dt^{2}$ are shown in Figs.~\ref{fig:betaRindler}, \ref{fig:fig_x}, \ref{fig:fig_p}, and \ref{fig:fig_a}, respectively.

% In Figs.~\ref{fig:fig_x} and \ref{fig:fig_a}, we see that antimatter follows the same dynamics as ordinary matter under the influence of gravity, thus numerically confirming the equivalence principle for Dirac anti-fermions.

In the momentum Ehrenfest theorem \eqref{ehren p}, the dependence on $\braket{\beta}$ superficially seems to suggest that the equivalence principle is violated. This conundrum is compounded when considering that the sign of $\braket{\beta}$ for matter and antimatter is positive and negative, respectively. However, our numerical simulations reveal that this is not the case. In Fig.~\ref{fig:betaRindler}, we see that $\braket{\beta}$ rapidly vanishes for both matter and antimatter, thus the dependence on $\braket{\beta}$ in Eq.~\eqref{ehren p} is effectively negligible especially given the short time scales. Additionally, in Figs.~\ref{fig:fig_x} and \ref{fig:fig_a}, we see that both matter and antimatter follow the same dynamics under the influence of gravity, thus obeying the equivalence principle.

To fully demystify the results of the mixed energy states in our simulations, we delve into the internal dynamics of the Dirac spinor itself. It is known that during time evolution, the internal degrees of freedom, i.e. matter and antimatter components, of a Dirac spinor can lead to nontrivial physical consequences \cite{fabbri_covariant_2018,fabbri_polar_2020,rogerio_general_2022}. The quantity that encodes this form of internal dynamics is the Yvon-Takabayashi  angle $\theta$ \cite{yvon_equations_1940,takabayasi_relativistic_1957} which is defined using the pseudoscalar $\Theta$ and scalar $\Phi$ bilinear covariant quantities
\begin{gather}
    \Theta = i\overline{\Psi}\gamma_{5}\Psi = 2\varphi^{2} \sin{\theta}, \quad \Phi = \overline{\Psi} \Psi = 2\varphi^{2} \cos{\theta}, \\
    \theta = \tan^{-1} \left( \frac{\Theta}{\Phi} \right),
\end{gather}
where the module $\varphi$ measures the density of the material distribution \cite{fabbri_generally_2016,fabbri_torsion_2018,fabbri_geometry_2019}. Rather than an explicit calculation, we can infer the value of the Yvon-Takabayashi angle for free-falling Dirac fermions via the Ehrenfest theorem of $\gamma_{5}$ which is
\begin{align}
    \frac{d}{dt} \braket{\gamma_{5}} &= \frac{1}{i\hbar} \braket{[\gamma_{5},\hat{H}_{g} ]} \notag \\
    &= \frac{2m}{i\hbar} \left(\braket{c^{2} \gamma_{5}\beta} + \braket{\gamma_{5}\beta g\hat{z}} \right), \label{ehrenfest y5}
\end{align}
where we used
\begin{align}
    [\gamma_{5},\beta] = 2\gamma_{5}\beta, \quad [\gamma_{5},\alpha_{i}] = 0.
\end{align}
Since $\braket{\gamma_{5}}$ is not conserved, the Yvon-Takabayashi angle is nonzero thus the spinor will undergo jittering motion or \emph{zitterbewegung}, the interference of the positive and negative energy states, even in the rest frame \cite{fabbri_geometry_2019}. The \emph{zitterbewegung} time scale is $t_\text{zitt} \sim \hbar / (2mc^2)$, which in the adopted natural units $t_\text{zitt} \sim 1/2$. We note that $t_\text{zitt}$ is the shortest time interval for which the single-particle interpretaion of the Diract equation is valid since the corresponding uncertainty in energy $\hbar / t_\text{zitt} \sim 2mc^2$ is sufficient to create an electron-positron pair, thereby  entering into the realm of quantum electrodynamics.

Figures~\ref{fig:fig_x} and \ref{fig:fig_a} show that the position and acceleration rapidly fluctuates at the onset of free fall. These oscillations look as if the gravity and anti-gravity are interchanging. Such transient effects are due to \emph{zitterbewegung} because of the time scale and the fact that the oscillations have the largest magnitude for the mixed energy state for which the particle-antiparticle interference is the strongest. We would like to name this observation as \textit{zitterbewegung-induced anti-gravity}. However, Figs.~\ref{fig:fig_x} and \ref{fig:fig_a} confirm that for longer non-transient times both matter and antimatter obey the equivalence principle.

\section{Entropic Gravity For Dirac Fermions} \label{section: relativistic dfeg master equation}

In this section, we formulate the DFEG Lindblad master equation for Dirac fermions. We use the formalism of open quantum systems and reservoir engineering \cite{vuglar_non-conservative_2018_v2} to construct a reservoir that simulates entropic gravity. We demonstrate that our DFEG model for Dirac fermions is decoherence-free and provide a physical analysis of our results. In addition, we find that entropic gravity does not affect spin. The results of this section will be crucial in developing the results in Sec.~\ref{section: relativistic qbounce hamiltonian}. 

Let $\hat{\rho}$ be the density matrix that represents the state of a mixture of Dirac fermions. A free-falling Dirac fermion in a linear gravitational potential 
\begin{align}
    V(\hat{z})=\beta mg\hat{z}, \label{relativistic linear potential}
\end{align}
is then described by the Liouville equation
\begin{align}
    \frac{d\hat{\rho}}{dt}  &= -\frac{i}{\hbar}\left[\hat{H}_{S} + \beta mg \hat{z}, \hat{\rho}\right], \label{linear gravity master eq} \\
    \hat{H}_{S} &=c\alpha_{3} \hat{p}_{3} + \beta mc^{2}, \label{system ham} 
\end{align} 
and its dynamics follow the free-fall Ehrenfest theorems \eqref{ehren x}-\eqref{ehren p} which can be shown using Eq.~\eqref{linear gravity master eq} and the density matrix expectation value
\begin{align}
    \langle \hat{O} \rangle = \text{Tr}\left[\hat{O}\hat{\rho} \right], \label{density matrix expect}
\end{align}
where $\hat{O}\equiv O(\hat{z},\hat{p}_{3})$ is an arbitrary observable. Eq.~\eqref{linear gravity master eq} is the conservative gravity master equation for Dirac fermions. Quantum coherence is encapsulated by the purity $\text{Tr}[\hat{\rho}^{2}]$, which is being conserved by  Eq.~\eqref{linear gravity master eq}.

There exists an infinite number of master equations that satisfy Eqs.~\eqref{ehren x}-\eqref{ehren p}. This means that we can find a master equation that mimics conservative gravity by utilizing a dissipator instead of using a potential. By carefully engineering an environment, a quantum system can obey the dynamics governed by the Ehrenfest-like equations \cite{schimmoller_decoherence-free_2020}
\begin{align}
    \frac{d}{dt} \braket{\hat{z}} = \braket{A(\hat{p}_{3})}, \quad \frac{d}{dt} \braket{\hat{p}_{3}} = \braket{B(\hat{z})}. 
\end{align}
For our purpose of modeling entropic gravity, we wish to engineer an environment that simulates the linear gravitational potential \eqref{relativistic linear potential} and follows the dynamics according to Eqs.~\eqref{ehren x}-\eqref{ehren p}. To achieve that, the paradigm of \textit{Operational Dynamical Modeling} (ODM) \cite{ODM-Bondar_2012} for spin-$1/2$ relativistic particles \cite{cabrera_operational_2019} is to be employed.

We use the simplest scenario of a single dissipator coupled to the closed system of a free Dirac fermion. In the formalism of open quantum systems theory, the Dirac fermion's density matrix $\hat{\rho}$ evolves according to the Lindblad master equation \cite{theory_of_open_quantum_systems_breuer}
\begin{align}
    \frac{d\hat{\rho}}{dt}  &= -\frac{i}{\hbar}\left[\hat{H}_{S},\hat{\rho} \right] + \mathcal{D} [\hat{\rho}], \label{density equations} \\
    \mathcal{D}[\hat{\rho}] &= \frac{\sigma}{\hbar} \left(\hat{A}\hat{\rho}\hat{A}^{\dagger} - \frac{1}{2} \{ \hat{A}^{\dagger}\hat{A}, \hat{\rho} \} \right). 
\end{align}
where the free parameter $\sigma \geq 0$ is the dimensionless coupling constant that quantifies the coupling strength and $\hat{A} \equiv A(\hat{z},\hat{p}_{3})$ is the unknown jump operator \cite{gallis_models_1993, vuglar_non-conservative_2018_v2,wigner-lindblad-eqs-quantum-friction}. To find the correct choice of $\hat{A}$ such that master equation \eqref{density equations} simulates an entropic gravity environment, we insert $\hat{z}$ and $\hat{p}_{3}$ into the Ehrenfest equation
\begin{align}
    \frac{d}{dt} \langle \hat{O} \rangle &= \frac{i}{\hbar} \Braket{\left[\hat{H}_{S},\hat{O}\right]} + \langle \mathcal{D}^{\dagger}[\hat{O}] \rangle, \notag \\
    \langle \mathcal{D}^{\dagger}[\hat{O}] \rangle &= \frac{\sigma}{\hbar} \text{Tr} \left[\left(\hat{A}^{\dagger}\hat{O}\hat{A} - \frac{1}{2} \{ \hat{A}^{\dagger}\hat{A}, \hat{O} \} \right) \hat{\rho} \right], \label{open quantum ehrenfest}
\end{align}
and set them equal to Eqs.~\eqref{ehren x} and \eqref{ehren p}, respectively, which yields 
\begin{align}
    \text{Tr}\left[\left(\hat{A}^{\dagger} \frac{\partial \hat{A}}{\partial \hat{p}_{3}} - \frac{\partial \hat{A}^{\dagger}}{\partial \hat{p}_{3}} \hat{A} \right) \hat{\rho}\right] &= 0, \label{trace a eq 1} \\
    \text{Tr}\left[\left(\hat{A}^{\dagger} \frac{\partial \hat{A}}{\partial \hat{z}} - \frac{\partial \hat{A}^{\dagger}}{\partial \hat{z}} \hat{A} \right)\hat{\rho}\right] &= -\frac{2img}{\sigma} \text{Tr}\left[\beta \hat{\rho}\right], \label{trace a eq 2}
\end{align}
where we have used the cyclic invariance property of the trace operation and
\begin{align}
    [A(\hat{z},\hat{p}_{3}),\hat{z}] = -i\hbar \frac{\partial\hat{A}}{\partial\hat{p}_{3}},  \quad [A(\hat{z},\hat{p}_{3}),\hat{p}_{3}] = i\hbar \frac{\partial\hat{A}}{\partial\hat{z}}. 
\end{align}
We demand that identities Eqs.~\eqref{trace a eq 1}-\eqref{trace a eq 2} hold for any arbitrary initial state, thus we drop the averaging and get the following constraint equations for $\hat{A}$
\begin{align}
    \hat{A}^{\dagger} \frac{\partial \hat{A}}{\partial \hat{p}_{3}} - \frac{\partial \hat{A}^{\dagger}}{\partial \hat{p}_{3}} \hat{A} &= 0, \label{A eq 1} \\
    \hat{A}^{\dagger} \frac{\partial \hat{A}}{\partial \hat{z}} - \frac{\partial \hat{A}^{\dagger}}{\partial \hat{z}} \hat{A} &= -\frac{2img}{\sigma} \beta. \label{A eq 2}
\end{align}
Any jump operator $\hat{A}$ that satisfies Eqs.~\eqref{A eq 1}-\eqref{A eq 2} will yield Eqs.~\eqref{ehren x}-\eqref{ehren p} when using Eq.~\eqref{density equations}, thus creating an entropic gravity environment that simulates the free-fall dynamics of conservative gravity.

Our choice of $\hat{A}$ is narrowed by the fact that $\hat{A}$ is not unique. Other than satisfying Eqs.~\eqref{A eq 1}-\eqref{A eq 2}, the choice of jump operator $\hat{A}$ must yield a master equation that is translationally invariant. This would make the master equation obey the strong equivalence principle \cite{di_casola_nonequivalence_2015} since the dynamics induced by the homogeneous gravitational field are independent of the choice of origin. According to Refs.~\cite{holevo_translation-covariant_1995,holevo_covariant_1996,petruccione_quantum_2005,vacchini_master-equations_2005,vacchini_quantum_2009,zhdanov_no_2017}, the following form is guaranteed to be translationary invariant
\begin{align}
    A(\hat{z},\hat{p}_{3}) &= e^{-iC\hat{z}} h(\hat{p}_{3}), \quad C^{\dagger} = C, \label{ansatz odm}
\end{align}
where $C$ is a Hermitian matrix. Inserting the ansatz \eqref{ansatz odm} into Eqs.~\eqref{A eq 1}-\eqref{A eq 2} yields 
\begin{align}
    \hat{h}^{\dagger} \frac{d \hat{h}}{d \hat{p}_{3}} - \frac{d \hat{h}^{\dagger}}{d \hat{p}_{3}} \hat{h} &= 0, \label{f eq 1} \\
    \hat{h}^{\dagger} C \hat{h} &= \frac{mg}{\sigma} \beta. \label{f eq 2}
\end{align}
There are many ways to satisfy Eqs.~\eqref{f eq 1}-\eqref{f eq 2} but for our work, we choose $h(\hat{p}_{3})=\sqrt{mgx_{0}} \mathbbm{1}_{4}$ such that $\hat{h}$ is a constant matrix. The value of the characteristic length $x_{0}$ will be determined in Sec.~\ref{section: nonrelativistic limit}. Then the Hermitian matrix \eqref{f eq 2} is
\begin{align}
    C = \frac{1}{x_{0} \sigma} \beta,
\end{align}
and the jump operator is
\begin{align}
    A(\hat{z},\hat{p}_{3}) &\equiv A(\hat{z}) = \sqrt{mg x_{0}} e^{-i\beta \hat{z}/(x_{0} \sigma)}. \label{bath A}
\end{align}
Finally, the DFEG equation for Dirac fermions is 
\begin{align}
    \frac{d\hat{\rho}}{dt}  &= -\frac{i}{\hbar} \left[c \alpha_{3} \hat{p}_{3} + \beta mc^{2},\hat{\rho}\right] + \mathcal{D} [\hat{\rho}], \label{relativistic dfeg master eq} \\
    \mathcal{D}[\hat{\rho}] &= \frac{mgx_{0}\sigma}{\hbar} \left(e^{-i\beta \hat{z}/(x_{0} \sigma)} \hat{\rho} e^{+i\beta \hat{z}/(x_{0} \sigma)} - \hat{\rho}\right). \label{rindler dissipator}
\end{align}
We shall refer to master equation \eqref{relativistic dfeg master eq} as the Dirac DFEG master equation or model. As mentioned before, the Dirac DFEG model \eqref{relativistic dfeg master eq} is translationally invariant hence it obeys the strong equivalence principle. 

To verify that our Dirac DFEG model \eqref{relativistic dfeg master eq} reproduces the conservative gravity model \eqref{linear gravity master eq} in the strong coupling limit $\sigma \rightarrow \infty$, we expand the exponential in dissipator \eqref{rindler dissipator} using the Baker-Campbell-Hausdorff (BCH) formula in the limit $\sigma \rightarrow \infty$ to get
\begin{align}
    \frac{d\hat{\rho}}{dt} &= -\frac{i}{\hbar}\left[c \alpha_{3} \hat{p}_{3} + \beta mc^{2} +\beta mg\hat{z},\hat{\rho}\right] \notag \\
    &+ \frac{mg}{x_{0} \hbar\sigma} \left(\hat{z}\beta\hat{\rho}\beta\hat{z} - \frac{1}{2} \{\hat{z}^{2},\hat{\rho} \} \right) + O(\sigma^{-2}). \label{master eq expansion}
\end{align}
The $O(\sigma^{-2})$ term in Eq.~\eqref{master eq expansion} quickly vanishes thus our Dirac DFEG model \eqref{relativistic dfeg master eq} reproduces the dynamics of a Dirac fermion subject to a linear gravitational potential, namely, master equation \eqref{linear gravity master eq}. 

%The $O(\sigma^{-2})$ term in Eq.~\eqref{master eq expansion} quickly vanishes thus in the strong coupling limit, our Dirac DFEG model \eqref{relativistic dfeg master eq} reproduces the dynamics of a Dirac fermion subject to a linear gravitational potential, namely, Eq.~\eqref{linear gravity master eq}. 

As proved in Appendix \ref{appendix: decoherence property}, the purity equation for the Dirac DFEG model in the strong coupling limit \eqref{master eq expansion} is
\begin{align}
    \frac{d}{dt} \text{Tr}[\hat{\rho}^{2}] &= -\frac{2mg}{x_{0} \hbar\sigma} \text{Tr}\left[\hat{\rho}^{2} \hat{z}^{2} -(\hat{\rho}\beta \hat{z})^{2}\right] + O(\sigma^{-2}), \label{purity change}
\end{align}
with $\text{Tr}[\hat{\rho}^{2} \hat{z}^{2} -(\hat{\rho}\beta \hat{z})^{2}] \geq 0$. Thus Eq.~\eqref{purity change} monotonically decreases as $\sigma \rightarrow \infty$ and larger $\sigma$ values, i.e. stronger coupling, can be chosen to preserve more purity leading to quantum coherence being maintained. Therefore, our Dirac DFEG model lives up to its namesake and is decoherence-free thus the argument that entropic gravity destroys quantum coherence is refuted for Dirac fermions. 

%with $\text{Tr}[\hat{\rho}^{2} \hat{z}^{2} -(\hat{\rho}\beta \hat{z})^{2}] \geq 0$. Thus Eq.~\eqref{purity change} monotonically decreases as $\sigma \rightarrow \infty$ and larger $\sigma$ values, i.e. stronger coupling, preserve more purity and a sufficiently large value of $\sigma$ can be chosen such that quantum coherence is maintained. Therefore, our Dirac DFEG model is decoherence-free and the argument that entropic gravity destroys quantum coherence is refuted for Dirac fermions. 

If we insert the low energy Hamiltonian \eqref{low energy rind ham} into the Dirac DFEG equation in the strong coupling limit \eqref{master eq expansion}, we get
\begin{align}
    \frac{d}{dt}\braket{\hat{H}_{g}} &= -\frac{2mgc}{x_{0}\hbar \sigma}  \Braket{\alpha_{3}  \hat{z}\hat{p}_{3}\hat{z}}+ O(\sigma^{-2}). \label{energy rate}
\end{align}
We see that the expected energy rate is dependent on the state due to the position and momentum operators present in \eqref{energy rate}. This is in stark contrast to the nonrelativistic DFEG model's constant expected rate of energy change $g\hbar/(2x_{0}\sigma)$ \cite[Eq.~(24)]{schimmoller_decoherence-free_2020}. 

To see the spin dynamics in our entropic gravity model \eqref{relativistic dfeg master eq}, we calculate its Ehrenfest theorem. 
% Recall that the $4\times 4$ spin observables are
% \begin{align}
%     \mathbf{S} &= \frac{\hbar}{2} \bm{\Sigma} = \frac{\hbar}{2}\begin{pmatrix}
%         \bm{\sigma} & 0 \\
%         0 & \bm{\sigma}
%     \end{pmatrix}, \label{spin matrix}
% \end{align}
% which have the commutation relations
% \begin{align}
%     [\alpha_{i},S_{j}] = i\hbar \varepsilon_{ijk} \alpha_{k}, \quad [\beta,S_{j}]=0, \label{spin comm rel}
% \end{align}
% where $\varepsilon_{ijk}$ is the Levi-Civita tensor. The first commutation relation in Eqs.~\eqref{spin comm rel} can be deduced using 
% \begin{align}
%     [\sigma^{i},\sigma^{j}] &= 2i \varepsilon^{ijk} \sigma^{k}.
% \end{align}
We see that inserting the spin \eqref{spin matrix} into dissipator \eqref{rindler dissipator} yields
\begin{align}
    \mathcal{D}[S_{j}] = \mathcal{D}^{\dagger} [S_{j}] =0, \label{spin diss}
\end{align}
where we have used Eqs.~\eqref{spin comm rel}, so the spin Ehrenfest theorems are
\begin{align}
    \frac{d}{dt} \Braket{S_{j}} &= \frac{i}{\hbar} \Braket{\left[c \alpha_{3} \hat{p}_{3} + \beta mc^{2},S_{j}\right]} + \Braket{\mathcal{D}^{\dagger} [S_{j}]} \notag \\
    &=  -\Braket{\varepsilon_{3jk} c\alpha_{k}\hat{p}_{3}},
\end{align}
or explicitly,
\begin{gather}
    \frac{d}{dt} \Braket{S_{1}} = -\Braket{c\alpha_{2}\hat{p}_{3}}, \quad \frac{d}{dt} \Braket{S_{2}} =  \Braket{c\alpha_{1}\hat{p}_{3}}, \label{ehrenfest spin x and y} \\
    \frac{d}{dt} \Braket{S_{3}} = 0. \label{ehrenfest spin z}
\end{gather}
We see from Eqs.~\eqref{spin diss} and \eqref{ehrenfest spin z} that the dissipator \eqref{rindler dissipator} preserves the spin along the direction of the acceleration. Thus our entropic gravity model implies that the free-fall dynamics of spin-$1/2$ Dirac fermions are the same as spinless particles therefore our model does not conflict with the equivalence principle. This is in agreement with recent experiments of the equivalence principle on spin-1/2 atoms which have demonstrated that an atom's spin \cite{tarallo_test_2014} and its orientation \cite{duan_test_2016} does not affect its free-fall dynamics, based on current sensitivity levels. 

We end this section by pointing out that while relativistic dissipative phenomena exist, there currently does not exist an accepted relativistic Lindblad master equation. D\'{i}osi \cite{diosi_is_2022} argued that it might be impossible to construct a relativistic Lindblad master equation. Nevertheless, there have been several proposals, such as covariant density matrix formulations \cite{gonzalez_relativistic_2022,cabrera_dirac_2016}, which have seen varying degrees of success in formulating relativistic open quantum systems. Since we are working in the low energy regime by using Hamiltonian \eqref{low energy rind ham}, our Dirac DFEG model \eqref{relativistic dfeg master eq} is a low energy model and is within the range of applicability of nonrelativistic open quantum systems. Therefore, our model avoids the ambiguity of relativistic open quantum systems. We will see later in Sec.~\ref{section: nonrelativistic limit} that our model \eqref{relativistic dfeg master eq} successfully reduces to the nonrelativistic DFEG model which makes our master equation \eqref{relativistic dfeg master eq} a satisfactory, ad-hoc model for Dirac fermions. 

%We end this section by making note of the fact that while relativistic dissipative phenomena are physically real, there currently does not exist an accepted relativistic Lindblad master equation. Some physicists  \cite{diosi_is_2022} even argue that it may be impossible to construct a relativistic Lindblad master equation. Nevertheless, there have been several proposals, such as covariant density matrix formulations \cite{gonzalez_relativistic_2022,cabrera_dirac_2016}, which have seen varying degrees of success in formulating relativistic open quantum systems. Since we are working in the low energy regime by using Hamiltonian \eqref{low energy rind ham}, our Dirac DFEG model \eqref{relativistic dfeg master eq} is a low energy model and is within the range of applicability of open quantum systems. Therefore, our model avoids the ambiguity of relativistic open quantum systems. We will see later in Sec.~\ref{section: nonrelativistic limit} that our Dirac DFEG model \eqref{relativistic dfeg master eq} successfully reduces to the nonrelativistic DFEG model which makes our master equation \eqref{relativistic dfeg master eq} a satisfactory, ad-hoc model for Dirac fermions. 

\section{Boundary Condition of Bouncing Dirac Fermion} \label{section: boundary condition of bouncing dirac fermion}

In this section, we choose the boundary condition from Ref.~\cite{boulanger_bound_2006} and present its re-derivation, which will be crucial in modeling the relativistic \qb experiment in Sec.~\ref{section: relativistic qbounce hamiltonian}. We give the rationale for such a choice along with some brief insight into the difficulty of imposing boundary conditions on the Dirac equation. 

The boundary condition, modeling the vibrating mirror in the \qb experiment, must satisfy two criteria if we are to relativistically generalize the \qb Hamiltonian in \cite{schimmoller_decoherence-free_2020}: $(1)$ the (relativistic) boundary condition should reduce to $J^{3}\lvert_{u=u_{0}}=0$, i.e., a vanishing probability current at the mirror's location $u_{0}$ ensuring that the Dirac fermions are reflected after hitting the mirror; and $(2)$ the (relativistic) boundary condition should reduce to the Dirichlet condition in the nonrelativistic limit.

Choosing appropriate boundary conditions that satisfy the criteria is rather complicated. The most logical choice is to utilize the Dirichlet condition, as is done for the nonrelativistic linear gravitational potential \cite{sakurai} and the Rindler space Klein-Gordon equation \cite{saa_bound_1997}, but imposing the Dirichlet condition on the Rindler Hamiltonian's \eqref{rind ham op form} eigenspinors (Eq.~\eqref{spin sol} in Appendix \ref{appendix: rindler dirac equation: eigenfunctions and eigenenergies}) leads to the trivial solution when calculating the energy levels \cite{rohim_relativistic_2021, boulanger_bound_2006}, which is clearly undesirable. Treading around this problem and directly using the vanishing probability current condition $J^{3} \lvert_{u=u_{0}} =0$ on the eigenspinors leads to the trivial solution as well \cite{rohim_relativistic_2021, boulanger_bound_2006}. The commonly used MIT boundary condition \cite{chodos_new_1974} satisfies our first criterion and also leads to energy quantization for the Rindler Hamiltonian \eqref{rind ham op form} (see Ref.~\cite{rohim_relativistic_2021} and Appendix \ref{appendix: rindler dirac equation: eigenfunctions and eigenenergies} for the energy levels) but does not reduce to the Dirichlet condition in the nonrelativistic limit \cite{alonso_general_1997}. Thus we rule out the MIT condition.

%Choosing appropriate boundary conditions satisfies our criteria turns out to be rather complicated. The most logical choice is to utilize the Dirichlet condition as is done for the nonrelativistic linear gravitational potential \cite{sakurai} and the Rindler space Klein-Gordon equation \cite{saa_bound_1997}, but imposing the Dirichlet condition on the wave functions (Eq.~\eqref{spin sol} in Appendix \ref{appendix: rindler dirac equation: eigenfunctions and eigenenergies}) of the Rindler Hamiltonian \eqref{rind ham op form} leads to the trivial solution $\Psi_{\Omega,s}=0$ when calculating the energy levels \cite{rohim_relativistic_2021, boulanger_bound_2006}, which is clearly undesirable. Treading around this problem and directly using the vanishing probability current condition $J^{3} \lvert_{u=u_{0}} =0$ on the Rindler Hamiltonian's wave functions also leads to the trivial solution \cite{rohim_relativistic_2021, boulanger_bound_2006}. One commonly used boundary condition is the MIT boundary condition \cite{chodos_new_1974} which satisfies our first criterion and leads to energy quantization for the Rindler Hamiltonian \eqref{rind ham op form} (see Ref.~\cite{rohim_relativistic_2021} and Appendix \ref{appendix: rindler dirac equation: eigenfunctions and eigenenergies} for the energy levels) but does not reduce to the Dirichlet condition in the nonrelativistic limit \cite{alonso_general_1997}. Thus we rule out using the MIT condition.

To avoid these issues and satisfy our criteria, we elect to utilize the boundary condition in \cite{boulanger_bound_2006} where we model the mirror as a scalar potential
\begin{align}
    \phi(z) &= 
    \begin{cases}
    0, & z>0, \\
    V_{0} \gg mc^{2}, & z\leq 0,
    \end{cases} \label{boulanger mirror potential}
\end{align}
or in Rindler coordinates,
\begin{align}
    \phi(u) &= 
    \begin{cases}
    0, & u>u_{0}, \\
    V_{0} \gg mc^{2}, & u\leq u_{0}, \label{boulanger mirror potential rindler}
    \end{cases} 
\end{align}
where $u_{0}=c^{2}/g$ is the mirror's location. Incorporating potential \eqref{boulanger mirror potential} essentially amounts to the replacement
\begin{align}
    mc^{2} \rightarrow \overline{m}c^{2} = mc^{2} + V_{0}, \label{mass replacement}
\end{align}
at the level of the Lagrangian \cite{boulanger_bound_2006} leading to the Rindler Hamiltonian \eqref{3d dirac ham op} to instead use the mass term \eqref{mass replacement}. We now closely follow Ref.~\cite{boulanger_bound_2006} and rederive the boundary condition that we will impose on the spinors of the Rindler Hamiltonian \eqref{3d dirac ham op}.

%We now closely follow Ref.~\cite{boulanger_bound_2006} and rederive the boundary condition that we will impose on the wave functions $\Psi$ of Rindler Hamiltonian \eqref{3d dirac ham op}. We begin with the derivation of a modified form of the Rindler Hamiltonian \eqref{3d dirac ham op} from which we will find the boundary condition on $\Psi$. For any spin-dependent spinor $\Psi_{s}$ that satisfies the Rindler Hamiltonian \eqref{3d dirac ham op}, we use the modified plane wave ansatz from Ref.~\cite{boulanger_bound_2006} which is

We begin with the derivation of a modified form of the Rindler Hamiltonian \eqref{3d dirac ham op} from which we find the differential equation for the spinor. For any positive frequency, spin-dependent spinor $\Psi_{s}$ that satisfies Hamiltonian \eqref{3d dirac ham op}, we use the modified plane wave ansatz \cite{boulanger_bound_2006}
\begin{align}
    \Psi_{s}(x) &= e^{-i\Omega v} e^{ikx} u^{-1/2} \psi_{s}(u), \quad \Omega>0, \label{boulanger plane wave sol} \\
    x &\equiv x^{\mu} = (v,x^{1},x^{2},u), \notag \\
    \mathbf{x} \equiv x^{a} &= (x^{1}, x^{2}, u), \quad a=1,2, \notag \\
    kx &\equiv \bm{k}_{\perp} \cdot \mathbf{x} = -k_{a} x^{a} = k^{1}x^{1} + k^{2} x^{2}, \notag \\
    k &= \abs{\bm{k}_{\perp}} = \sqrt{(k^{1})^{2} + (k^{2})^{2}}, \notag 
\end{align}
where $\Omega = \omega c/g$ is the dimensionless frequency. Note that ansatz \eqref{boulanger plane wave sol} differs from the usual plane wave ansatz \eqref{plane wave sol} in Appendix \ref{appendix: rindler dirac equation: eigenfunctions and eigenenergies}. Inserting ansatz \eqref{boulanger plane wave sol} into Eq.~\eqref{3d cov rind dir eq} yields
\begin{align}
    \left[\Omega \gamma^{0} + \slashed{k}u + i u \gamma^{3}\partial_{u} - \overline{\ell} u \right] \psi_{s} = 0, \label{boulanger ansatz dirac eq}
\end{align}
where $\slashed{k}=k_{a}\gamma^{a}=-(k^{1}\gamma^{1} + k^{2} \gamma^{2})$ and $\overline{\ell}=\overline{m}c/\hbar$. Then we use another ansatz that decomposes the spatial spinor $\psi_{s}$ as 
\begin{align}
    \psi_{s}(u) &= F(u) U_{s} + G(u)\gamma^{3}U_{s}, \label{boulanger spinor ansatz} 
\end{align}
where the spin-dependent spinor $U_{s}$ obeys 
\begin{align}
    U^{\dagger}_{s} U_{s^{\prime}} = \delta_{s,s^{\prime}}, \quad
    \gamma^{0}U_{s} = U_{s}, \quad \hat{\slashed{k}}\gamma^{3}U_{s} = is U_{s}, \label{boulanger gamma0 U cond}
\end{align}
and $\hat{\slashed{k}}=\slashed{k}/k$ \cite{boulanger_bound_2006}. Note that Eqs.~\eqref{boulanger gamma0 U cond} also imply that 
\begin{align}
    U^{\dagger}_{s} \gamma^{3} U_{s} = 0.
\end{align}
Using the ansatz decomposition \eqref{boulanger spinor ansatz} on Eq.~\eqref{boulanger ansatz dirac eq} then decouples Eq.~\eqref{boulanger ansatz dirac eq} into
\begin{align}
    \begin{cases}
    \phantom{+} D_{-}F - iks G- i(\partial_{u}G)=0, \\
    -D_{+}G - iks F + i(\partial_{u}F)=0,
    \end{cases} \label{boulanger F and G eqs}
\end{align}
where we have used the linear independence of $U_{s}$ and $\gamma^{3}U_{s}$ to decouple $F$ and $G$ and
\begin{align}
    D_{\pm} &= \frac{\Omega}{u} \pm \overline{\ell}.
\end{align}
We can use the first and second equations from Eqs.~\eqref{boulanger F and G eqs} to get
\begin{align}
    F &= \frac{i}{D_{-}} (\partial_{u} + ks)G, \label{boulanger derivative f} \\
    G &= \frac{i}{D_{+}} (\partial_{u} - ks)F, \label{boulanger derivative g}
\end{align}
respectively, thus we can find $G$ from $F$ and vice versa. We choose to focus on $F$ and use identity \eqref{boulanger derivative g} on the first equation in Eqs.~\eqref{boulanger F and G eqs} to get 
\begin{align}
    \left[\partial^{2}_{u} - \frac{(\partial_{u}D_{+})}{D_{+}} (\partial_{u}- ks) +D_{+}D_{-} - k^{2}s^{2} \right] F =0,
\end{align}
which we simplify further using
\begin{align}
    F(u) &= D^{1/2}_{+}f(u), \label{really big f identity}
\end{align}
which leads to our desired differential equation
\begin{align}
    &\left[ \partial^{2}_{u} + D_{+}D_{-} - k^{2}s^{2} + \frac{(\partial_{u}D_{+})}{D_{+}}ks + \frac{(\partial^{2}_{u}D_{+})}{2D_{+}} \right. \notag \\
    &\left. - \frac{3(\partial_{u}D_{+})^{2}}{4D^{2}_{+}} \right] f = 0. \label{f dirac equation}
\end{align}

With Eq.~\eqref{f dirac equation}, we can now find the boundary condition for spinor \eqref{boulanger spinor ansatz} by analyzing the asymptotic behavior of the function $f(u)$ based on the mirror potential \eqref{boulanger mirror potential rindler}. In the region $u>u_{0}$, $\phi(u)=0$ so $\overline{m}=m$ and we are left with the Rindler Hamiltonian \eqref{3d dirac ham op} thus there are no continuity rules to impose on spinor \eqref{boulanger spinor ansatz}. In the region $u<u_{0}$, $\phi(u)=V_{0} \gg mc^{2}$ so if we take the limit $V_{0}\rightarrow \infty$, we have
\begin{align}
    D_{\pm} &\approx \pm \frac{V_{0}}{\hbar c},
\end{align}
which leads to Eq.~\eqref{f dirac equation} asymptotically reducing to
\begin{align}
    (\partial^{2}_{u}f) \approx \left(\frac{V_{0}}{\hbar c} \right)^{2} f. \label{infinite M limit}
\end{align}
Solving Eq.~\eqref{infinite M limit} yields
\begin{align}
    f(u) \approx N e^{V_{0}u/\hbar c},
\end{align}
where $N$ is a normalization constant. Then we use identities \eqref{really big f identity} and \eqref{boulanger derivative g} to find that
\begin{align}
    F(u) \approx N e^{V_{0}u/\hbar c}, \quad G(u) \approx iN e^{V_{0}u/\hbar c}, \label{boulanger F and G}
\end{align}
respectively, in the $u<u_{0}$ region. The wave function must be continuous at the mirror's location $u=u_{0}$ which we can impose by assuming from Eq.~\eqref{boulanger F and G} the Robin boundary condition
\begin{align}
    F \Big\lvert_{u=u_{0}} &= -iG \Big\lvert_{u=u_{0}}. \label{boulanger bc}
\end{align}
Thus we finally have our boundary condition for the wave function. To explicitly see why condition \eqref{boulanger bc} is a Robin condition, we use identities \eqref{boulanger derivative f}-\eqref{boulanger derivative g} to rewrite condition \eqref{boulanger bc} for $F$ and $G$ as
\begin{align}
    (D_{+} + ks )  F \Big\lvert_{u=u_{0}} &= (\partial_{u}F) \Big\lvert_{u=u_{0}}, \label{boulanger spin bc 2} \\
    -(D_{-} + ks )  G \Big\lvert_{u=u_{0}} &= (\partial_{u}G) \Big\lvert_{u=u_{0}}, \label{boulanger spin bc 1} 
\end{align}
respectively.

For subsequent sections, it would prove far more fruitful to have boundary condition \eqref{boulanger bc} in spinor form $\Psi_{s}$. To achieve this, we simply impose condition \eqref{boulanger bc} on the decomposition ansatz \eqref{boulanger spinor ansatz} and use Eqs.~\eqref{boulanger gamma0 U cond} to get
\begin{align}
    i\gamma^{3} \Psi_{s} \Big\lvert_{u=u_{0}} &= \Psi_{s} \Big\lvert_{u=u_{0}}, \label{boulanger spinor bc} 
\end{align}
which is the MIT boundary condition for chiral angle $\theta_{\text{M}}=0$ \cite{chodos_new_1974, nicolaevici_bouncing_2017}. The difference between the MIT condition and condition \eqref{boulanger spinor bc} is that the MIT condition focuses only on the spinor $\Psi_{s}$, making no explicit assumption as to the form of the spinor, while our condition \eqref{boulanger spinor bc} focuses on the functions $F$ and $G$ that the spinor $\Psi_{s}$ is composed of. 

We now verify that boundary condition \eqref{boulanger spinor bc} satisfies our criteria. To test the first criterion, we calculate the probability current which is
\begin{align}
    J^{3} \Big\lvert_{u=u_{0}} &= \overline{\Psi}_{s} \gamma^{3} \Psi_{s} \Big\lvert_{u=u_{0}} \notag \\
    &=\frac{i}{D_{+}}(F(\partial_{u}F)^{\ast} - F^{\ast}(\partial_{u}F)) \Big\lvert_{u=u_{0}}. \label{boulanger vanishing current F}
\end{align}
Now $F$ and $(\partial_{u}F)$ are determined up to an arbitrary complex coefficient thus we choose them to be real functions which make the probability current \eqref{boulanger vanishing current F} vanish at the boundary \cite{boulanger_bound_2006}
\begin{align}
    J^{3} \Big\lvert_{u=u_{0}} &= \overline{\Psi}_{s} \gamma^{3} \Psi_{s} \Big\lvert_{u=u_{0}} =0.\label{boulanger vanishing current}
\end{align}
We can also prove Eq.~\eqref{boulanger vanishing current} by using condition \eqref{boulanger spinor bc}, and its adjoint form
\begin{align}
    -i\overline{\Psi}_{s}\gamma^{3} \Big\lvert_{u=u_{0}} &= \overline{\Psi}_{s} \Big\lvert_{u=u_{0}},
\end{align}
to get \cite{rohim_relativistic_2021}
\begin{align}
    J^{3} \Big\lvert_{u=u_{0}} &= \overline{\Psi}_{s} \gamma^{3} \Psi_{s} \Big\lvert_{u=u_{0}} \notag \\
    &= \overline{\Psi}_{s} \Psi_{s} \Big\lvert_{u=u_{0}}= -\overline{\Psi}_{s} \Psi_{s} \Big\lvert_{u=u_{0}}= 0. 
\end{align}
Thus our first criterion is satisfied. In the nonrelativistic limit, we can expand the exponentials in Eqs.~\eqref{boulanger F and G} up to $O(c^{-1})$ then impose condition \eqref{boulanger bc} to get 
\begin{align}
    F \Big\lvert_{u=u_{0}} \approx N = -iN \approx G \Big\lvert_{u=u_{0}},
\end{align}
which implies that \cite{boulanger_bound_2006}
\begin{align}
    F \Big\lvert_{u=u_{0}} \approx 0, \quad G \Big\lvert_{u=u_{0}} \approx 0. \label{boulanger non rel f and g}
\end{align}
In spinor form, conditions \eqref{boulanger non rel f and g} trivially lead to
\begin{align}
    \Psi_{s} \Big\lvert_{u=u_{0}} \approx 0, \label{nonrel boulanger cond}
\end{align}
thus boundary condition \eqref{boulanger bc} nonrelativistically reduces to the Dirichlet condition, satisfying our second criterion. 

We now derive a crucial identity that will be used in the next section. We take the first derivative of ansatz \eqref{boulanger spinor ansatz} and impose the boundary condition \eqref{boulanger bc} to get
\begin{align}
    (\partial_{u}\psi_{s}) \Big\lvert_{u=u_{0}} &= (\partial_{u} F) U_{s} + (\partial_{u}G)\gamma^{3}U_{s} \Big\lvert_{u=u_{0}} \notag \\
    &= \Omega_{u}\left(F U_{s} - G\gamma^{3}U_{s} \right)+ \ell \psi_{s} \Big\lvert_{u=u_{0}} \notag \\
    &= \Omega_{u}\left(F \gamma^{0} U_{s} - G\gamma^{3}(\gamma^{0})^{2}U_{s} \right) + \ell \psi_{s} \Big\lvert_{u=u_{0}} \notag \\
    &= \left(\Omega_{u}\gamma^{0}+ \ell \right) \psi_{s} \Big\lvert_{u=u_{0}}, \\
    \Omega_{u} &= \left(\Omega u^{-1} + ks \right),
\end{align}
where we used identities \eqref{boulanger derivative f}-\eqref{boulanger derivative g}, $(\gamma^{0})^{2}=1$, and \eqref{boulanger gamma0 U cond} in the second, third, and fourth equalities, respectively. Then we have
\begin{align}
    \overline{\psi}_{s}(\partial_{u}\psi_{s}) \Big\lvert_{u=u_{0}} &=  \overline{\psi}_{s} \left(\Omega_{u}\gamma^{0}+ \ell \right) \psi_{s} \Big\lvert_{u=u_{0}} \notag \\
    &= (\partial_{u}\overline{\psi}_{s})\psi_{s}  \Big\lvert_{u=u_{0}},
\end{align}
therefore
\begin{align}
    (\partial_{u}\overline{\Psi}_{s})\Psi_{s} \Big\lvert_{u=u_{0}} =\overline{\Psi}_{s} (\partial_{u}\Psi_{s})\Big\lvert_{u=u_{0}}, \label{boulanger derivative equivalence rindler}
\end{align}
or equivalently in the observer's coordinates,
\begin{align}
    (\partial_{3}\overline{\Psi}_{s})\Psi_{s} \Big\lvert_{z=0} =\overline{\Psi}_{s} (\partial_{3}\Psi_{s})\Big\lvert_{z=0}. \label{boulanger derivative equivalence}
\end{align}

\section{Relativistic \texorpdfstring{\MakeLowercase{\emph{q}}Bounce}{qBounce} Hamiltonian} \label{section: relativistic qbounce hamiltonian}

In this section, we relativistically model \qb experiment by using the boundary condition from Sec.~\ref{section: boundary condition of bouncing dirac fermion} to find the surface term that arises from the Rindler Hamiltonian's Ehrenfest theorems. Although we will ultimately use the low energy Hamiltonian \eqref{low energy rind ham}, we first use the linear high energy Rindler Hamiltonian \eqref{rind ham op form} to not miss any low-order terms. We will remove the high-order terms and modify the low energy Hamiltonian \eqref{low energy rind ham} with the surface term. Then, we will apply the results of Sec.~\ref{section: relativistic dfeg master equation} to formulate the relativistic master equations that reproduce the \qb experiment.

We follow the methodology in \cite{schimmoller_decoherence-free_2020} and modify the Ehrenfest theorems of the high-energy, linear Rindler Hamiltonian \eqref{rind ham op form}. We require a temporal variable which we easily identify from Eqs.~\eqref{rindler coordinates} as $v$, so the Ehrenfest theorems for the Rindler position and momentum operators \eqref{rind position momentum ops} are 
\begin{align}
    \frac{d}{dv} \langle \hat{u} \rangle = \langle \alpha_{3}  \hat{u} \rangle, \quad \frac{d}{dv} \langle \hat{p}_{u} \rangle = -\langle \alpha_{3}  \hat{p}_{u} \rangle - mc \langle \beta \rangle,
\end{align}
respectively. However, the above equations do not incorporate the contribution of boundary condition \eqref{boulanger bc}, i.e. the vibrating mirror, so we integrate by parts with the Dirac inner product \eqref{dirac inner product} to find the surface term generated by condition \eqref{boulanger bc}. Following this prescription for the Rindler momentum operator $\hat{p}_{u}$ yields
\begin{align}
    \frac{d}{dv} \langle \hat{p}_{u} \rangle &= -\langle \alpha_{3} \hat{p}_{u} \rangle - mc\langle \beta \rangle -\frac{i\hbar c^{2}}{g}\overline{\Psi}\gamma^{3} (\partial_{u} \Psi) \Big\lvert_{u=u_{0}} \notag \\
    &= -\langle \alpha_{3} \hat{p}_{u} \rangle - mc\langle \beta \rangle + \frac{\hbar c^{2}}{g}\overline{\Psi} (\partial_{u} \Psi) \Big\lvert_{u=u_{0}} \notag \\
    &= -\langle \alpha_{3} \hat{p}_{u} \rangle - mc\langle \beta \rangle + \Braket{f_{R}(\hat{u})}, \label{surface term derivation}
\end{align}
where we used condition \eqref{boulanger spinor bc} in the second equality and have identified the Rindler statistical force
\begin{align}
    \Braket{f_{R}(\hat{u})} &= \frac{\hbar c^{2}}{2g} \Braket{\beta \delta^{\prime}(\hat{u} - u_{0})}, \label{rindler surface term}
\end{align}
where $u_{0}=c^{2}/g$, as the surface term that arises from boundary condition \eqref{boulanger bc}. The $1/2$ factor comes from identity \eqref{boulanger derivative equivalence rindler} and the Dirac delta function $\delta(x)$ is defined as
\begin{align}
    \int^{\infty}_{-\infty} dx \, \delta^{(n)}(x-x^{\prime}) f(x) &= (-1)^{n} f^{(n)}(x^{\prime}).
\end{align}
Note that we have suppressed the spin subscript $s$. Similar calculations for the Rindler position operator $\hat{u}$ yields
\begin{align}
    \frac{d}{dv} \langle \hat{u} \rangle &= \langle \alpha_{3} \hat{u} \rangle + \left( \frac{c^{4}}{g^{2}} J^{3} \right) \Big\lvert_{u=u_{0}} = \langle \alpha_{3} \hat{u} \rangle,
\end{align}
where we used the vanishing probability current condition \eqref{boulanger vanishing current}. Thus the Ehrenfest theorems for the Rindler position and momentum operators with boundary condition \eqref{boulanger bc} are
\begin{alignat}{3}
    &\frac{d}{dv} \langle \hat{u} \rangle &&= \phantom{+} \langle \alpha_{3} \hat{u} \rangle, \label{bc rindler ehrenfest position} \\
    &\frac{d}{dv} \langle \hat{p}_{u} \rangle &&= -\langle \alpha_{3} \hat{p}_{u} \rangle - mc\langle \beta \rangle + \Braket{f_{R}(\hat{u})}, \label{bc rindler ehrenfest momentum} 
\end{alignat}
respectively. Using Eqs.~\eqref{rindler coordinates} on Eqs.~\eqref{bc rindler ehrenfest position}-\eqref{bc rindler ehrenfest momentum} yields the Ehrenfest theorems in the observer's local coordinates $(t,z)$ 
\begin{alignat}{3}
    &\frac{d}{dt} \langle \hat{z} \rangle &&= \phantom{+} \frac{g}{c}\langle \alpha_{3} \hat{z} \rangle + c\langle \alpha_{3} \rangle, \\
    &\frac{d}{dt} \langle \hat{p}_{3} \rangle &&= -\frac{g}{c}\langle \alpha_{3} \hat{p}_{3} \rangle - mg\langle \beta \rangle + \Braket{f_{S}(\hat{z})}, 
\end{alignat}
which, in the low energy regime, reduces to 
\begin{alignat}{3}
    &\frac{d}{dt} \langle \hat{z} \rangle &&= \phantom{+} c\langle \alpha_{3} \rangle, \label{coordinate ehrenfest position}  \\
    &\frac{d}{dt} \langle \hat{p}_{3} \rangle &&= - mg\langle \beta \rangle + \Braket{f_{S}(\hat{z})}, \label{coordinate ehrenfest momentum}
\end{alignat}
where 
\begin{align}
    \braket{f_{S}(\hat{z})} &= \frac{\hbar c}{2} \Braket{\beta \delta^{\prime}(\hat{z})}, \label{observer surface term}
\end{align}
is the surface term \eqref{rindler surface term} in the observer's coordinates.

We add the surface term \eqref{observer surface term} to the system Hamiltonian \eqref{system ham} to get the relativistic \qb Hamiltonian 
\begin{align}
    \hat{H}_{q} &=c\alpha_{3} \hat{p}_{3} + \beta mc^{2}+ V_{q}(\hat{z}) - \frac{\hbar c}{2} \beta \delta(\hat{z}), \label{low energy oscillating ham}
\end{align}
where 
\begin{align}
    V_{q}(\hat{z}) &= \begin{cases}
        \beta mg\hat{z}, \quad &\text{conservative gravity}, \\
        0, \quad &\text{entropic gravity}.
    \end{cases}
\end{align}
To make the boundary oscillate, we add a sinusoidal term in the argument of the delta function as follows
\begin{align}
    \hat{H}_{q} &=c\alpha_{3} \hat{p}_{3} + \beta mc^{2}+ V_{q}(\hat{z}) -\frac{\hbar c}{2} \beta \delta(\hat{z} - a_{m}\sin{\omega_{m} t}), \label{oscillating ham}
\end{align}
where $a_{m}$ and $\omega_{m}$ are the vibrating mirror's oscillation strength and frequency, respectively. We can now apply the Dirac DFEG model \eqref{relativistic dfeg master eq} to Hamiltonian \eqref{oscillating ham} to get the relativistic \qb experiment's master equations for conservative and entropic gravity 
\begin{align}
    \frac{d\hat{\rho}}{dt}  &= -\frac{i}{\hbar} \left[ \vphantom{\frac{1}{1}} c \alpha_{3} \hat{p}_{3} + \beta mc^{2}+\beta mg\hat{z} \right. \notag \\
    &\left. -\frac{\hbar c}{2} \beta \delta(\hat{z} - a_{m}\sin{\omega_{m} t}),\hat{\rho} \right], \label{qbounce master eq} \\
    \frac{d\hat{\rho}}{dt}  &= -\frac{i}{\hbar} \left[ \vphantom{\frac{1}{1}} c \alpha_{3} \hat{p}_{3} + \beta mc^{2} \right. \notag \\
    &\left. -\frac{\hbar c}{2} \beta \delta(\hat{z} - a_{m}\sin{\omega_{m} t}),\hat{\rho} \right] + \mathcal{D} [\hat{\rho}], \label{entropic master eq} 
\end{align}
respectively.

Compared to the \qb Hamiltonian's boundary term in \cite[Eq.~(16)]{schimmoller_decoherence-free_2020} which was proportional to the first derivative of the Dirac delta function, the relativistic Hamiltonian's \eqref{oscillating ham} boundary term is $\propto$ the Dirac delta function. In addition, the appearance of the $\beta$ matrix captures the effect of the mirror on both matter and antimatter as well. 

%In particular, the relativistic boundary term switches signs for antimatter meaning that antiparticles exhibit the opposite behavior due to the vibrating mirror. 
%We discuss this further in Sec.~\ref{section: nonrelativistic limit}.

\section{Nonrelativistic Limit} \label{section: nonrelativistic limit}

With our master equations fully developed, we now present the nonrelativistic limit of our results. We use the Foldy-Wouthysen (FW) transformation \cite{foldy_dirac_1950} to find the nonrelativistic approximation of our Hamiltonians then use the FW Hamiltonians to construct their corresponding nonrelativistic master equations. 

\subsection{Hamiltonian}

We begin with a brief overview of the FW transformation. Let $\Psi(\mathbf{x},t)$ be an arbitrary Dirac spinor that satisfies the general Dirac equation
\begin{align}
    i\hbar \partial_{t}\Psi &= \hat{H}_{\mathbf{D}}\Psi, \label{general dirac eq} \\
    \hat{H}_{\mathbf{D}} &= c\bm{\alpha} \cdot \hat{\mathbf{p}} + \beta mc^{2} + V_{D}(\hat{\mathbf{x}}) + K(\hat{\mathbf{p}}), \label{general dirac ham}
\end{align}
where $V_{D}(\hat{\mathbf{x}})$ and $K(\hat{\mathbf{p}})$ are matrix-valued functions of operators $\hat{\mathbf{x}}$ and $\hat{\mathbf{p}}$, respectively. Following the standard convention, we define the even and odd components of Hamiltonian \eqref{general dirac ham} as \cite{foldy_dirac_1950,silenko_foldywouthuysen_2003}
\begin{align}
    \hat{\mathcal{E}} = \frac{1}{2}(\hat{H}_{\mathbf{D}} + \beta \hat{H}_{\mathbf{D}} \beta), \quad \hat{\mathcal{O}} =\frac{1}{2}(\hat{H}_{\mathbf{D}} - \beta \hat{H}_{\mathbf{D}} \beta),
\end{align}
respectively, where 
\begin{align}
    [\beta,\hat{\mathcal{E}}] =\{\beta,\hat{\mathcal{O}} \} = 0. \label{even odd comms}
\end{align}
We also define the following operators
\begin{align}
    \hat{S}_{1} = -\frac{i\beta \hat{\mathcal{O}}}{2mc^{2}} , \quad \hat{U}_{1} = e^{i\hat{S}_{1}},
\end{align}
which are Hermitian and unitary, respectively, where the subscript denotes the number of FW transformations applied. Then Hamiltonian \eqref{general dirac ham} can be written as 
\begin{equation}
    \hat{H}_{\mathbf{D}} = \hat{\mathcal{O}} + \hat{\mathcal{E}}, \label{general dirac ham with e and o}
\end{equation}
and applying the first FW transformation on Hamiltonian \eqref{general dirac ham with e and o} yields \cite{foldy_dirac_1950}
\begin{align}
    \hat{H}_{1} &= \hat{U}_{1}\hat{H}_{\mathbf{D}}\hat{U}^{\dagger}_{1}, \label{first fw trans} \\
    \Psi_{1} &= \hat{U}_{1} \Psi, 
\end{align}
which turns Eq.~\eqref{general dirac eq} into
\begin{align}
    i\hbar \partial_{t} \Psi_{1} &= \hat{H}_{1} \Psi_{1}.
\end{align}
We can then evaluate Eq.~\eqref{first fw trans} up to a desired order via the BCH expansion and subsequent FW transformations can be performed using
\begin{align}
    \hat{S}_{n} = -\frac{i\beta\hat{\mathcal{O}}_{n-1}}{2mc^{2}}, \quad \hat{U}_{n} = e^{i\hat{S}_{n}}, \label{general unitary fw op}
\end{align}
which yields
\begin{align}
    i\hbar \partial_{t} \Psi_{FW} &= \hat{H}_{FW} \Psi_{FW}, \\
    \hat{H}_{FW} &\equiv \hat{H}_{n} = \hat{U}_{n}\hat{H}_{n-1}\hat{U}^{\dagger}_{n}, \\
    \Psi_{FW} &\equiv \Psi_{n} = \hat{U}_{n}\Psi_{n-1},
\end{align}
where $n=1,2,\ldots$. For most of our Hamiltonians, we require three FW transformations thus applying the FW transformation three times to remove all odd operators yields \cite{hehl_inertial_1990}
\begin{align}
    \hat{H}_{FW} &\equiv \hat{H}_{3} = \beta \left(mc^{2}+  \frac{\hat{\mathcal{O}}^{2}}{2mc^{2}} - \frac{\hat{\mathcal{O}}^{4}}{8m^{3}c^{6}} \right) + \hat{\mathcal{E}} \notag \\
    &- \frac{1}{8m^{2}c^{4}}[\hat{\mathcal{O}},[\hat{\mathcal{O}},\hat{\mathcal{E}}]], \label{three time fw formula}
\end{align}
which turns Eq.~\eqref{general dirac eq} into
\begin{align}
    i\hbar \partial_{t} \Psi_{FW} &= \hat{H}_{FW} \Psi_{FW}, \label{fw dirac eq} \\
    \Psi_{FW} &= \Psi_{3} \equiv \hat{U}_{3} \Psi_{2},
\end{align}
where we have replaced the subscripts with $FW$ for clarity. 

For the system Hamiltonian \eqref{system ham}, a single FW transformation is sufficient and yields \cite{foldy_dirac_1950}
\begin{align}
    \hat{H}_{FW,S} \equiv \hat{H}_{1,S} &= \beta \sqrt{c^{2} \hat{p}^{2}_{3}  + m^{2}c^{4}} \approx \beta \left(mc^{2} + \frac{\hat{p}^{2}_{3}}{2m} \right),
\end{align}
where we have dropped terms of order $O(c^{-2})$ and higher. By using Eq.~\eqref{three time fw formula}, the low energy Hamiltonian's FW version is 
\begin{align}
    \hat{H}_{FW,\mathbf{g}} &= \beta \left(mc^{2} + \frac{\hat{\mathbf{p}}^{2}}{2m} - \frac{\hat{\mathbf{p}}^{4}}{8m^{3}c^{2}} + m(\mathbf{a} \cdot \hat{\mathbf{x}}) \notag  \right. \\
    &\left. +\frac{\hbar}{2mc^{2}} \bm{\Sigma} \cdot (\mathbf{a} \times \hat{\mathbf{p}}) \right), \label{nonrelativistic 3d rind ham} \\
    \mathbf{a} &= (0,0,g).
\end{align}
Note that we have used the three spatial dimensional version of the low energy Hamiltonian \eqref{low energy rind ham} to incorporate the spin contribution. The linear case with no high-energy corrections is
\begin{align}
    \hat{H}_{FW,g} &= \beta \left( mc^{2} + \frac{\hat{p}^{2}_{3}}{2m} + mg \hat{z} \right). \label{nonrelativistic rind ham}
\end{align}

Reducing the relativistic \qb Hamiltonian \eqref{low energy oscillating ham} to its nonrelativistic limit requires a rather different approach due to the nature of the surface term \eqref{rindler surface term}. Recall that the surface term is dependent on the choice of the boundary condition and since our surface term \eqref{rindler surface term} was derived using the \textit{relativistic} boundary condition \eqref{boulanger bc} and the Rindler Hamiltonian \eqref{rind ham op form}, our term \eqref{rindler surface term} is inherently \textit{relativistic}. Naively applying the FW transformation on the three spatial dimensional version of Hamiltonian \eqref{low energy oscillating ham} would yield (for $V_{\mathbf{q}}(\hat{z})=\beta m(\mathbf{a}\cdot\hat{\mathbf{x}})$)
\begin{align}
    \hat{H}_{FW,\mathbf{q}} &= \beta \left(mc^{2} + \frac{\hat{\mathbf{p}}^{2}}{2m} - \frac{\hat{\mathbf{p}}^{4}}{8m^{3}c^{2}} + m(\mathbf{a} \cdot \hat{\mathbf{x}}) -\frac{\hbar c}{2} \delta(\hat{\mathbf{x}}) \right. \notag \\
    &\left. + \frac{\hbar}{2mc^{2}} \bm{\Sigma} \cdot (\mathbf{a} \times \hat{\mathbf{p}}) \right), \label{wrong fw oscillating ham}
\end{align}
but we have only transformed the Hamiltonian while the boundary term
\begin{align}
    -\frac{\hbar c}{2} \beta \delta(\hat{\mathbf{x}}), \label{dubious surface term}
\end{align}
is \textit{still relativistic} since we have not changed the boundary condition. Also, we should be able to reproduce the boundary term \eqref{dubious surface term} from the momentum Ehrenfest theorem of Hamiltonian \eqref{nonrelativistic 3d rind ham} using the \textit{nonrelativistic} limit of boundary condition \eqref{boulanger bc}, i.e. the Dirichlet condition \eqref{nonrel boulanger cond}, but a quick calculation will show that this is not the case. Thus, these issues force us to rule out FW Hamiltonian \eqref{wrong fw oscillating ham} as the correct nonrelativistic version of Hamiltonian \eqref{low energy oscillating ham}.

To resolve these problems and find the FW form of Hamiltonian \eqref{low energy oscillating ham}, we follow the same procedure in deriving surface term \eqref{rindler surface term} except we must use the FW Hamiltonian \eqref{nonrelativistic rind ham} and the nonrelativistic Dirichlet condition \eqref{nonrel boulanger cond}. Thus we take inspiration from the Heisenberg and Schrödinger pictures from nonrelativistic quantum mechanics and use the FW unitary operator \eqref{general unitary fw op} to define the FW picture. We will first derive the surface term with conservative gravity and then generalize to include the entropic case. Let $\ket{\Psi(t)}$ and $\hat{Q}$ be a general Dirac four-component state and observable, respectively, then the expectation value can be expressed as
\begin{align}
    \langle \hat{Q} \rangle &= \bra{\Psi} \hat{U}^{\dagger}_{n} \hat{U}_{n} \hat{Q} \hat{U}^{\dagger}_{n} \hat{U}_{n} \ket{\Psi} \notag \\
    &= \bra{\Psi_{FW}} \hat{Q}_{FW} \ket{\Psi_{FW}} \notag \\
    &= \langle \hat{Q}_{FW} \rangle_{FW},
\end{align}
where the averaging is now taken with respect to the FW Dirac state $\ket{\Psi_{FW}} = \hat{U}_{n}\ket{\Psi}$. Then an operator in the FW picture is \cite{foldy_dirac_1950,nieto_hamiltonian_1977}
\begin{align}
    \hat{Q}_{FW} &= \hat{U}^{\dagger}_{n} \hat{Q} \hat{U}_{n},
\end{align}
and its equation of motion is
\begin{align}
    \frac{d\hat{Q}_{FW}}{dt} &= \frac{i}{\hbar} [\hat{H}_{FW},\hat{Q}_{FW}].
\end{align}
The FW operator $\hat{Q}_{FW}$ can be evaluated using the BCH expansion in a similar fashion to an FW Hamiltonian, but such an expansion will often have a complicated expression due to the form of $\hat{S}_{n}$ and the number of times an FW transformation has been applied. However, if $\hat{Q}$ is of order $O(c^{0})$ or less, we can use the BCH expansion up to the first order 
\begin{align}
    \hat{Q}_{FW} &= e^{i\hat{S}_{n}} \hat{Q} e^{-i\hat{S}_{n}} \notag \\
    &= \hat{Q} +i[\hat{S}_{n},\hat{Q}] + \frac{i^{2}}{2!} [\hat{S}_{n},[\hat{S}_{n},\hat{Q}]] + \ldots \notag \\
    &\approx \hat{Q},
\end{align}
as a sufficient approximation since the lowest order term in $\hat{S}_{n}$ is $O(c^{-1})$. Analogously, we must use the same cutoff order $O(c^{0})$ for the FW Hamiltonian if we are to maintain symmetry in calculating the Ehrenfest theorems.

For our work, we use $\hat{U}_{3}$ so the approximated FW position and momentum operators are 
\begin{align}
    \hat{z}_{FW} \approx \hat{z}, \quad \hat{p}_{3,FW} \approx \hat{p}_{3},
\end{align}
respectively, with their FW Ehrenfest theorems being
\begin{align}
    &\frac{d}{dt} \Braket{\hat{z}_{FW}}_{FW} \approx \frac{d}{dt} \Braket{\hat{z}}_{FW} \notag \\
    &\qquad = \frac{1}{m} \Braket{\beta \hat{p}_{3}}_{FW} -\frac{i\hbar}{2m} \overline{\Psi}_{FW} \Psi_{FW} \Big\lvert_{z=0}, \label{fw position ehrenfest derivation} \\
    &\frac{d}{dt} \Braket{\hat{p}_{3,FW}}_{FW} \approx \frac{d}{dt} \Braket{\hat{p}_{3}}_{FW} \notag \\
    &\qquad= -mg\Braket{\beta}_{FW} - \frac{\hbar^{2}}{2m} \overline{\Psi}_{FW} (\partial^{2}_{3} \Psi_{FW}) \Big\lvert_{z=0} \notag \\
    &\qquad + \frac{\hbar^{2}}{2m} (\partial_{3} \overline{\Psi}_{FW}) (\partial_{3}\Psi_{FW}) \Big\lvert_{z=0}, \label{fw momentum ehrenfest derivation}
\end{align}
where we used Eq.~\eqref{fw dirac eq} and FW Hamiltonian \eqref{nonrelativistic rind ham}. To evaluate the surface terms in Eqs.~\eqref{fw position ehrenfest derivation}-\eqref{fw momentum ehrenfest derivation}, recall that the boundary condition \eqref{boulanger bc} nonrelativistically reduced to the Dirichlet condition \eqref{nonrel boulanger cond} so the boundary condition for the FW spinor is simply
\begin{align}
    \Psi_{FW} \Big\lvert_{u=u_{0}} \approx 0, \label{nonrel fw bc}
\end{align}
so we get
\begin{alignat}{3}
    &\frac{d}{dt} \Braket{\hat{z}}_{FW} &&= \phantom{+} \frac{1}{m} \Braket{\beta \hat{p}_{3}}_{FW}, \\
    &\frac{d}{dt} \Braket{\hat{p}_{3}}_{FW} &&= -mg\Braket{\beta}_{FW} + \frac{\hbar^{2}}{4m} \Braket{\beta \delta^{\prime \prime}(\hat{z})}_{FW}.
\end{alignat}
Then the FW \qb Hamiltonian with conservative gravity is
\begin{align}
    \hat{H}_{FW,q} &= \beta \left( mc^{2} + \frac{\hat{p}^{2}_{3}}{2m} + mg\hat{z} - \frac{\hbar^{2}}{4m} \delta^{\prime}(\hat{z}) \right), \label{nonrel qbounce ham}
\end{align}
and more generally, 
\begin{align}
    \hat{H}_{FW,q} &= \beta \left( mc^{2} + \frac{\hat{p}^{2}_{3}}{2m} + V_{FW,q}(\hat{z}) - \frac{\hbar^{2}}{4m} \delta^{\prime}(\hat{z}) \right), \label{nonrel qbounce ham general}
\end{align}
where we have included the entropic case by using 
\begin{align}
    V_{FW,q}(\hat{z}) &= \begin{cases}
        mg\hat{z}, \quad &\text{conservative gravity}, \\
        0, \quad &\text{entropic gravity}.
    \end{cases}
\end{align}
Without the rest energy, Hamiltonian \eqref{nonrel qbounce ham general} is
\begin{align}
    \hat{H}_{FW,q} &= \beta \left(\frac{\hat{p}^{2}_{3}}{2m} + V_{FW,q}(\hat{z}) - \frac{\hbar^{2}}{4m} \delta^{\prime}(\hat{z}) \right),
\end{align}
which is the \qb Hamiltonian in \cite[Eq.~(15)]{schimmoller_decoherence-free_2020}. Since the boundary term in \eqref{nonrel qbounce ham} was derived using the nonrelativistic condition \eqref{nonrel boulanger cond} and \eqref{nonrelativistic rind ham}, Hamiltonian \eqref{nonrel qbounce ham} is mathematically and physically symmetric. Thus we can definitively interpret FW Hamiltonian \eqref{nonrel qbounce ham} as the nonrelativistic approximation of \eqref{low energy oscillating ham}. 

%which is the \qb Hamiltonian in \cite{schimmoller_decoherence-free_2020}. Since the surface term in FW Hamiltonian \eqref{nonrel qbounce ham} was derived with the FW Ehrenfest theorem and cut off the same order of terms as in Hamiltonian \eqref{nonrelativistic rind ham}, Hamiltonian \eqref{nonrel qbounce ham} is mathematically and physically symmetric. In addition, the surface term in Hamiltonian \eqref{nonrel qbounce ham} was derived from the nonrelativistic, Dirichlet boundary condition \eqref{nonrel boulanger cond} so we can definitively interpret the surface term and FW Hamiltonian \eqref{nonrel qbounce ham} as the nonrelativistic approximation of the relativistic \qb Hamiltonian \eqref{low energy oscillating ham}. 

%Thus we see that utilizing the FW picture successfully reduced our relativistic \qb Hamiltonian to the \qb Hamiltonian in \cite{schimmoller_decoherence-free_2020}. 

% We again observe that anti-gravity enters the antimatter component in \eqref{nonrel qbounce ham}. In addition, the sign shift in the $\beta$ matrix causes the mirror to shift up the energy of the anti-fermion while the ordinary fermion experiences an energy decrease from the mirror.

\subsection{Master Equation}

To find the nonrelativistic limit of master equations \eqref{relativistic dfeg master eq}, \eqref{qbounce master eq}, and \eqref{entropic master eq}, we first find the nonrelativistic limit of the jump operator $\hat{A}$. Since the $\beta$ matrix is diagonal, the jump operator $\hat{A}$ can be analytically evaluated to get
\begin{align}
    \hat{A} &= \sqrt{mg x_{0}} e^{-i\beta \hat{z}/(x_{0} \sigma)} \notag \\
    &= \sqrt{mg x_{0}} \begin{pmatrix}
    e^{-i\hat{z}/(x_{0}\sigma)} \mathbbm{1}_{2} & 0 \\
    0 & e^{+i\hat{z}/(x_{0}\sigma)} \mathbbm{1}_{2}
    \end{pmatrix},
\end{align}
then dissipator \eqref{rindler dissipator} in the nonrelativistic limit is simply
\begin{align}
    \mathcal{D}_{NR}[\hat{\rho}] &= \frac{mgx_{0}\sigma}{\hbar} \left(e^{-i\beta \hat{z}/(x_{0} \sigma)} \hat{\rho} e^{+i\beta \hat{z}/(x_{0} \sigma)} - \hat{\rho} \right) \notag \\
    &= \frac{mgx_{0}\sigma}{\hbar} \left( R_{1} e^{-i\hat{z}/(x_{0}\sigma)} \hat{\rho} e^{+i\hat{z}/(x_{0}\sigma)} \right. \notag \\
    &\left. + R_{2} e^{+i\hat{z}/(x_{0}\sigma)} \hat{\rho} e^{-i\hat{z}/(x_{0}\sigma)} - \mathbbm{1}_{4} \hat{\rho} \right), \label{nonrelativistic dissipator}
\end{align}
where
\begin{align}
    R_{1} = \begin{pmatrix}
    \mathbbm{1}_{2} & 0 \\
    0 & 0
    \end{pmatrix}, \quad 
    R_{2} = \begin{pmatrix}
    0 & 0 \\
    0 & \mathbbm{1}_{2}
    \end{pmatrix}.
\end{align}
In a more compact form, dissipator \eqref{nonrelativistic dissipator} in component form is
\begin{align}
    \mathcal{D}_{\pm}[\hat{\rho}_{\pm}] &= \frac{mgx_{0}\sigma}{\hbar} \left(e^{\mp i \hat{z}/(x_{0} \sigma)} \hat{\rho}_{\pm} e^{\pm i \hat{z}/(x_{0} \sigma)} - \hat{\rho}_{\pm}\right), \label{nonrelativistic dissipator component}
\end{align}
where the sign subscript denotes the positive and negative frequency components of the density matrix. Then master equations \eqref{relativistic dfeg master eq}, \eqref{qbounce master eq}, and \eqref{entropic master eq} in the nonrelativistic limit are
\begin{align}
    \frac{d\hat{\rho}}{dt} &= -\frac{i}{\hbar} \left[\beta \left( mc^{2} + \frac{\hat{\mathbf{p}}^{2}}{2m} \right), \hat{\rho} \right] + \mathcal{D}_{NR}[\hat{\rho}], \label{nonrel rindler master eq} \\
    \frac{d\hat{\rho}}{dt}  &= -\frac{i}{\hbar} \left[\hat{H}_{FW,g}-\beta \frac{\hbar^{2}}{4m} \delta^{\prime}(\hat{\mathbf{x}} - a_{m}\sin{\omega_{m} t}),\hat{\rho} \right], \label{nonrel qbounce} \\
    \frac{d\hat{\rho}}{dt}  &= -\frac{i}{\hbar} \left[\beta \left( mc^{2} + \frac{\hat{\mathbf{p}}^{2}}{2m} - \frac{\hbar^{2}}{4m} \delta^{\prime}(\hat{\mathbf{x}} - a_{m}\sin{\omega_{m} t})  \right),\hat{\rho} \right] \notag \\
    &+\mathcal{D}_{NR}[\hat{\rho}], \label{nonrel dfeg with qbounce}
\end{align}
respectively, where we have used the three spatial dimensional Hamiltonians to incorporate the spin. For the positive frequency linear case with no rest energies and $O(c^{-2})$ terms, we get
\begin{align}
    \frac{d\hat{\rho}_{+}}{dt} &= - \frac{i}{\hbar} \left[\frac{\hat{p}^{2}_{3}}{2m}, \hat{\rho}_{+} \right] + \mathcal{D}_{+}[\hat{\rho}_{+}], \label{alex entropic master eq} \\
    \frac{d\hat{\rho}_{+}}{dt}  &= -\frac{i}{\hbar} \left[\frac{\hat{p}^{2}_{3}}{2m} + mg \hat{z} -\frac{\hbar^{2}}{4m} \delta^{\prime}(\hat{z} - a_{m}\sin{\omega_{m} t}),\hat{\rho}_{+} \right], \\
     \frac{d\hat{\rho}_{+}}{dt}  &= -\frac{i}{\hbar} \left[\frac{\hat{p}^{2}_{3}}{2m} -\frac{\hbar^{2}}{4m} \delta^{\prime}(\hat{z} - a_{m}\sin{\omega_{m} t}),\hat{\rho}_{+} \right] \notag \\
     &+ \mathcal{D}_{+}[\hat{\rho}_{+}].
\end{align}
which are the DFEG, conservative gravity, and entropic master equations \cite[Eqs.~(5), (17) and 
 (18)]{schimmoller_decoherence-free_2020}, respectively. Thus we conclude that the Dirac DFEG model \eqref{relativistic dfeg master eq} is the appropriate relativistic generalization, for Dirac fermions, of the nonrelativistic DFEG model. We are then able to identify the characteristic length $x_{0}$ as the same characteristic length value used in \cite[Eq.~(7)]{schimmoller_decoherence-free_2020}, namely, 
\begin{align}
    x_{0} = \left( \frac{\hbar^{2}} {2m^{2} g} \right)^{1/3}. \label{x0 value}
\end{align}
Therefore, our Dirac DFEG model \eqref{relativistic dfeg master eq} is physical.

\section{Discussion and Outlook} \label{section: discussion and outlook}

We have presented a generalized version of the DFEG model for Dirac fermions via the open quantum systems approach. In addition, we have presented a relativistic model of the \qb experiment with conservative and entropic gravity. In the nonrelativistic limit, our Dirac DFEG \eqref{relativistic dfeg master eq} and \qb models \eqref{qbounce master eq}-\eqref{entropic master eq} correctly reduced to their nonrelativistic counterparts in \cite{schimmoller_decoherence-free_2020}.

%We have shown that the derived Dirac DFEG model \eqref{relativistic dfeg master eq} maintains the quantum purity of a Dirac fermion even in the strong coupling limit. This means that in an entropic theory of gravity, the quantum coherence of a Dirac fermion is maintained even when strongly coupled $\sigma \rightarrow \infty$ to gravity. Also, we have shown that conservative gravity for Dirac fermions \eqref{linear gravity master eq} is reproduced in the strong coupling limit $\sigma \rightarrow \infty$. We see that spin is not affected by our entropic gravity model thus entropic gravity does not conflict with the equivalence principle. Thus, we have refuted entropic gravity's decoherence argument for Dirac fermions. 

We have shown that the derived Dirac DFEG model \eqref{relativistic dfeg master eq} maintains the quantum purity of a Dirac fermion in the strong coupling limit $\sigma \rightarrow \infty$. In the same limit, we have shown that conservative gravity \eqref{linear gravity master eq} for Dirac fermions is reproduced as well. Our model predicts that a Dirac fermion's spin does not affect its free-fall dynamics nor couple with gravity therefore our model does not conflict with the equivalence principle. Thus, we have refuted entropic gravity's decoherence argument for Dirac fermions and demonstrated that entropic gravity is compatible with conservative gravity.

%We have shown that the derived Dirac DFEG model \eqref{relativistic dfeg master eq} maintains the quantum purity of a Dirac fermion even in the strong coupling limit. In a bigger picture, since the Dirac equation describes Dirac fermions, our Dirac DFEG model \eqref{relativistic dfeg master eq} shows that in an entropic theory of gravity, the quantum coherence of a Dirac fermion is maintained even when strongly coupled $\sigma \rightarrow \infty$ to gravity. Also, we have shown that conservative gravity for Dirac fermions \eqref{linear gravity master eq} is reproduced in the strong coupling limit $\sigma \rightarrow \infty$. We see that spin is not affected by our entropic gravity model thus entropic gravity does not conflict with the equivalence principle. Thus, we have refuted entropic gravity's decoherence argument for Dirac fermions. 

From numerical simulations of Hamiltonian \eqref{low energy rind ham} and its Ehrenfest theorems \eqref{ehren x}-\eqref{ehren p}, we demonstrated that antimatter obeys the equivalence principle. In addition, we numerically found that the nonzero Yvon-Takabayashi angle led to a transient zitterbewegung-induced anti-gravity effect during the early stages of a mixed energy state's time evolution. The already ephemeral zitterbewegung-induced anti-gravity effect quickly diminishes for larger $g$ values thus we concluded that mixed energy states obeyed the equivalence principle as well.

We aim to conduct numerical simulations of the Dirac DFEG model to see how it compares with its nonrelativistic counterpart. Although the nonrelativistic \qb experiment \cite{cronenberg_acoustic_2018} is the best for measuring neutron free fall, our relativistic \qb model \eqref{entropic master eq} may potentially provide further refinements to the value of $\sigma$. Recent proposals \cite{gurlebeck_boost_2018,schiller_space_2017,noauthor_zarm_nodate} and developments \cite{sorrentino_space_2011,frye_bose-einstein_2021,devani_gravity_2020} in next-generation, space-based quantum experiments will potentially provide experimental data to test our work in the near future (see Ref.~\cite{belenchia_quantum_2022} for a thorough review). We hope that our work, backed by new data, will shed further light on whether gravity is truly quantum or not and spark further research into alternative theories of gravity such as entropic gravity.

\acknowledgments

D.I.B. was financially supported by the W. M. Keck Foundation and by Army Research Office (ARO) (grant W911NF-23-1-0288, program manager Dr.~James Joseph, and cooperative agreement W911NF-21-2-0139). E.J.S. was supported by the ARO Undergraduate Research Apprenticeship Program. The views and conclusions contained in this document are those of the authors and should not be interpreted as representing the official policies, either expressed or implied, of ARO or the U.S. Government. The U.S. Government is authorized to reproduce and distribute reprints for Government purposes notwithstanding any copyright notation herein.

\textbf{Note: When the current paper was accepted, an experimental study was published ruling out antigravity~\cite{anderson_observation_2023}. This verifies the conclusion of Sec.~\ref{section: antigravity and ehrenfest theorems} that both matter and antimatter obey the equivalence principle. We further note that the results of Ref.~\cite{anderson_observation_2023} say nothing against the transient phenomenon of \emph{zitterbewegung} antigravity predicted in Sec.~\ref{section: antigravity and ehrenfest theorems}.}

\appendix
\renewcommand{\theequation}{\thesection\arabic{equation}}

\section{Decoherence-Free Property of Dirac DFEG Model} \label{appendix: decoherence property}

In this section, we closely follow Ref.~\cite{vuglar_non-conservative_2018_v2} and show that the Dirac DFEG model \eqref{relativistic dfeg master eq} is decoherence-free in the strong coupling limit $\sigma \rightarrow \infty$ by proving that 
\begin{align}
    \text{Tr}\left[\hat{\rho}^{2}\hat{z}^{2} - (\hat{\rho}\beta \hat{z})^{2}\right] \geq 0,
\end{align}
in the purity equation \eqref{purity change}. 

We first expand the exponential term in jump operator \eqref{bath A} in the limit $\sigma \rightarrow \infty$ to get
\begin{align}
   e^{\pm i\beta\hat{z}/(x_{0}\sigma)} &= 1 \pm \frac{i\beta \hat{z}}{x_{0}\sigma} + \frac{1}{2!} \left( \frac{\pm i \beta \hat{z}}{x_{0}\sigma} \right)^{2} + O(\sigma^{-3}) \notag \\
    &=  \pm \frac{i\beta \hat{z}}{x_{0}\sigma} -\frac{\hat{z}^{2}}{2x^{2}_{0}\sigma^{2}} + O(\sigma^{-3}),
\end{align}
where we used both positive and negative signs for generality. Then in the limit $\sigma \rightarrow \infty$, master equation \eqref{relativistic dfeg master eq} is 
\begin{align}
    \frac{d\hat{\rho}}{dt}  &= -\frac{i}{\hbar} \left[c \alpha_{3} \hat{p}_{3} + \beta mc^{2} + \beta mg\hat{z},\hat{\rho} \right] \notag \\
    &+ \frac{mg}{x_{0}\hbar \sigma} \left( \beta \hat{z} \hat{\rho} \hat{z} \beta - \frac{1}{2} \{\hat{z}^{2},\hat{\rho} \} \right) + O(\sigma^{-2}), 
\end{align}
and the purity equation is
\begin{align}
    \frac{d}{dt} \text{Tr} \left[\hat{\rho}^{2} \right] &= -\frac{2mg}{x_{0} \hbar\sigma} \text{Tr}\left[\hat{\rho}^{2} \hat{z}^{2} -(\hat{\rho}\beta \hat{z})^{2} \right] + O(\sigma^{-2}).
\end{align}
Now for any two arbitrary operators $\hat{C}_{1}$ and $\hat{C}_{2}$, we have by the Cauchy-Schwartz inequality 
\begin{align}
    \text{Tr}\left[\hat{C}^{\dagger}_{1} \hat{C}_{1}\right] \, \text{Tr}\left[\hat{C}^{\dagger}_{2} \hat{C}_{2}\right] \geq \abs{\text{Tr}\left[\hat{C}^{\dagger}_{1} \hat{C}_{2} \right]}^{2},
\end{align}
so if we let $\hat{C}_{1} = \beta \hat{z} \hat{\rho}$ and $\hat{C}_{2}=\hat{\rho} \beta \hat{z}$, we get
\begin{align}
    \text{Tr}\left[\hat{\rho}^{2} \hat{z}^{2} \right] \geq \abs{\text{Tr}\left[(\hat{\rho} \beta \hat{z})^{2} \right]},
\end{align}
where we have used the cyclic property of the trace. Since $\text{Tr}[(\hat{\rho} \beta \hat{z})^{2} ]^{\ast} = \text{Tr}[(\hat{\rho} \beta \hat{z})^{2}]$, we have that $\text{Tr}[(\hat{\rho} \beta \hat{z})^{2}] \in \mathbb{R}$ thus we get our desired result
\begin{align}
    \text{Tr}\left[\hat{\rho}^{2}\hat{z}^{2} - (\hat{\rho}\beta \hat{z})^{2}\right] \geq 0.
\end{align}

\section{Rindler Dirac Equation: Eigenfunctions and Eigenenergies} \label{appendix: rindler dirac equation: eigenfunctions and eigenenergies}

In this section, we solve the full Rindler Hamiltonian \eqref{3d dirac ham op} by following Refs.~\cite{greiner_quantum_1985,rohim_relativistic_2021} (see also Refs.~\cite{soffel_dirac_1980,suzuki_analytic_2003,boulanger_bound_2006,crispino_unruh_2008} and Ref.~\cite{ueda_entanglement_2021} for a thorough review on solving the Rindler Dirac equation in all wedges of Minkowski space). Although we ultimately use the low energy Rindler Dirac Hamiltonian \eqref{low energy rind ham}, the exact solution and eigenenergies of the full Rindler Hamiltonian \eqref{3d dirac ham op} will prove fruitful to our later discussion. Note that solving the full Rindler Hamiltonian \eqref{3d dirac ham op} is equivalent to solving the full observer's Hamiltonian \eqref{3d non FW ham}.

A general Dirac wave packet in Rindler space is composed of positive and negative frequency states
\begin{align}
    \Psi(x) &= \int^{\infty}_{0} d\Omega \int^{\infty}_{-\infty} \frac{d^{2}\bm{k}_{\perp}}{2\pi} \sum_{s=\pm}  \left[ \vphantom{\left( \Psi^{+}_{\Omega,s,\bm{k}_{\perp}}(x) \right)^{C}} b(\Omega,s,\bm{k}_{\perp}) \Psi^{+}_{\Omega,s,\bm{k}_{\perp}}(x) \right. \notag \\
    &+ \left. d^{\ast}(\Omega,s,\bm{k}_{\perp}) \left( \Psi^{+}_{\Omega,s,\bm{k}_{\perp}}(x) \right)^{C} \right] \notag \\
    &= \int^{\infty}_{0} d\Omega \int^{\infty}_{-\infty} \frac{d^{2}\bm{k}_{\perp}}{2\pi} \sum_{s=\pm}  \left[ \vphantom{\Psi^{-}_{\Omega,-s,\bm{k}_{\perp}}(x)} b(\Omega,s,\bm{k}_{\perp}) \right. \Psi^{+}_{\Omega,s,\bm{k}_{\perp}}(x) \notag \\
    &+ \left. d^{\ast}(\Omega,s,\bm{k}_{\perp}) \Psi^{-}_{\Omega,-s,\bm{k}_{\perp}}(x) \right], \\
    d^{2} \bm{k}_{\perp} &\equiv dk^{1} \, dk^{2}, \notag
\end{align}
where $b(\Omega,s,\bm{k}_{\perp})$ and $d^{\ast}(\Omega,s,\bm{k}_{\perp})$ are the positive and negative energy wave amplitudes, respectively, and
\begin{align}
    &x \equiv x^{\mu} = (v,x^{1},x^{2},u), \notag \\
    &\mathbf{x} \equiv x^{a} = (x^{1},x^{2},u), \quad \bm{k}_{\perp} \equiv k^{a} = (k^{1},k^{2},0), \quad a = 1,2. \notag
\end{align}
The negative energy states are computed using the charge conjugation operator 
\begin{align}
    \left( \Psi^{+}_{\Omega, s, \bm{k}_{\perp}}(x) \right)^{C} &= i \gamma^{2} \left( \Psi^{+}_{\Omega,s,\bm{k}_{\perp}}(x) \right)^{\ast} = \Psi^{-}_{\Omega,-s,\bm{k}_{\perp}}(x).
\end{align}
We solve for the positive energy stationary states by using the plane wave ansatz to separate the temporal and spatial components of the positive energy Dirac spinor
\begin{align}
    \Psi^{+}_{\Omega,s,\bm{k}_{\perp}}(x) &= e^{-i\Omega v} f^{+}_{\Omega,s,\bm{k}_{\perp}}(\mathbf{x}), \quad \Omega>0, \label{plane wave sol} \\
    f^{+}_{\Omega,s,\bm{k}_{\perp}}(\mathbf{x}) &= e^{i kx } \psi^{+}_{\Omega,s,\bm{k}_{\perp}}(u), \\
    kx \equiv \bm{k}_{\perp} \cdot \mathbf{x} &= -k_{a}x^{a} = k^{1}x^{1} + k^{2} x^{2}, \notag \\
    k &= \abs{\bm{k}_{\perp}} = \sqrt{(k^{1})^{2} + (k^{2})^{2}}, \notag 
\end{align}
where $\Omega = \omega c/g$ is the dimensionless frequency. Note that we have separated the spatial components as well. Inserting ansatz \eqref{plane wave sol} into Eq.~\eqref{3d dirac ham} gives
\begin{align}
    \hbar \Omega \psi^{+}_{\Omega,s,\bm{k}_{\perp}} &= \left[\hbar u (\alpha_{1} k^{1} + \alpha_{2} k^{2} - i\alpha_{3} \partial_{u} ) -\frac{i\hbar}{2}\alpha_{3} \right. \notag \\
    &+\left. \vphantom{\frac{1}{1}} \beta mcu \right] \psi^{+}_{\Omega,s,\bm{k}_{\perp}}, 
\end{align}
which after rearranging terms yields 
\begin{align}
    &\left[u (\alpha_{1} k^{1} + \alpha_{2} k^{2} ) -i\alpha_{3} u\partial_{u} - \frac{i}{2}\alpha_{3} \right. \notag \\
    &+ \left.\vphantom{\frac{1}{2}} \beta \ell u - \Omega \right] \psi^{+}_{\Omega,s,\bm{k}_{\perp}} = 0, \label{sector I eq}
\end{align}
where $\ell =1/\lambdabar=mc/\hbar$ is the inverse reduced Compton wavelength. Next, we define the operator
\begin{align}
    D_{1} \equiv u \left(\alpha_{1} k^{1} + \alpha_{2} k^{2} \right) -i\alpha_{3} u\partial_{u} - \frac{i}{2}\alpha_{3} + \beta \ell u - \Omega,
\end{align}
then Eq.~\eqref{sector I eq} can be written as
\begin{align}
    D_{1}\psi^{+}_{\Omega,s,\bm{k}_{\perp}} = 0. \label{sector I eq operator}
\end{align}
We then define a similar operator 
\begin{align}
    D_{2} \equiv u \left(\alpha_{1} k^{1} + \alpha_{2} k^{2} \right) -i\alpha_{3} u\partial_{u} +\frac{i}{2}\alpha_{3} + \beta \ell u + \Omega,
\end{align}
and multiply $D_{2}$ on the left of Eq.~\eqref{sector I eq operator} to get
\begin{align}
    0 &= D_{2} D_{1} \psi^{+}_{\Omega,s,\bm{k}_{\perp}} \notag \\
    &= \left[-u \partial_{u}u\partial_{u} + \kappa^{2}u^{2} + \frac{1}{4} -\Omega^{2} - i\Omega \alpha_{3} \right] \psi^{+}_{\Omega,s,\bm{k}_{\perp}}, \notag
\end{align}
which yields after rearrangement
\begin{equation}
    u \partial_{u}u\partial_{u} \psi^{+}_{\Omega,s,\bm{k}_{\perp}} =\left[ \kappa^{2}u^{2} + \frac{1}{4} - \Omega^{2} - i\Omega \alpha_{3} \right] \psi^{+}_{\Omega,s,\bm{k}_{\perp}}, \label{final psi eq}
\end{equation}
where $\kappa = \sqrt{k^{2}+\ell^{2}}$. Now let $\chi_{1}(u)$ and $\chi_{2}(u)$ be the two-component spinors such that
\begin{equation}
    \psi^{+}_{\Omega,s,\bm{k}_{\perp}}(u) = \begin{pmatrix}
    \chi_{1}(u) \\
    \chi_{2}(u)
    \end{pmatrix}, \label{chi spinor}
\end{equation}
and inserting spinor \eqref{chi spinor} into Eq.~\eqref{final psi eq} yields the system of equations 
\begin{align}
     u \partial_{u}u\partial_{u} \chi_{1} &=\left( \kappa^{2}u^{2} + \frac{1}{4} -\Omega^{2} \right) \chi_{1} - i\Omega \sigma^{3} \chi_{2}, \label{coupled eq 1} \\
     u \partial_{u}u\partial_{u} \chi_{2} &=\left( \kappa^{2}u^{2} + \frac{1}{4} -\Omega^{2} \right) \chi_{2} - i\Omega \sigma^{3} \chi_{1}. \label{coupled eq 2}
\end{align}
Subtracting and adding Eqs.~\eqref{coupled eq 1} and \eqref{coupled eq 2} yields
\begin{align}
    u \partial_{u}u\partial_{u} (\chi_{1} - \chi_{2}) &=\left( \kappa^{2}u^{2} + \frac{1}{4} -\Omega^{2} \right) (\chi_{1} - \chi_{2}) \notag \\
    &+ i\Omega  \sigma^{3} (\chi_{1} - \chi_{2}), \\
     u \partial_{u} u \partial_{u} (\chi_{1} + \chi_{2})  &=\left( \kappa^{2}u^{2} + \frac{1}{4} -\Omega^{2} \right) (\chi_{1} + \chi_{2}) \notag \\
     &- i\Omega  \sigma^{3} (\chi_{1} + \chi_{2}),
\end{align}
which can be expressed in a more compact form as
\begin{align}
    u \partial_{u}u\partial_{u} (\chi_{1} \mp\chi_{2}) &=\left( \kappa^{2}u^{2} + \frac{1}{4} -\Omega^{2} \right) (\chi_{1} \mp \chi_{2}) \notag \\
    &\pm i\Omega \sigma^{3} (\chi_{1} \mp \chi_{2}).
\end{align}
To fully decouple Eqs.~\eqref{coupled eq 1} and \eqref{coupled eq 2}, we let
\begin{equation}
    \chi_{1}(u) \mp \chi_{2}(u) = \begin{pmatrix}
    \xi^{ \pm}(u) \\
    \zeta^{ \pm}(u)
    \end{pmatrix},
\end{equation}
which leads to
\begin{align}
    u \partial_{u}u\partial_{u} \xi^{ \pm} &=\left[ \kappa^{2}u^{2} + \left(i\Omega \mp \frac{1}{2}\right)^{2} \right]\xi^{ \pm}, \label{bessel eq 1} \\
    u \partial_{u}u\partial_{u} \zeta^{ \pm} &=\left[ \kappa^{2}u^{2} + \left(i\Omega \pm \frac{1}{2}\right)^{2} \right]\zeta^{ \pm}. \label{bessel eq 2}
\end{align}
These are Bessel's differential equations thus the spin state solutions are \cite{soffel_dirac_1980,greiner_quantum_1985,suzuki_analytic_2003}
\begin{align}
    \Psi^{+}_{\Omega,s,\bm{k}_{\perp}}(x) &= e^{-i\Omega v} f^{+}_{\Omega,s,\bm{k}_{\perp}}(\mathbf{x}), \label{full spin sol} \\
    f^{+}_{\Omega,s,\bm{k}_{\perp}}(\mathbf{x}) &= \mathcal{N}_{\Omega,\bm{k}_{\perp}} e^{i kx } \psi^{+}_{\Omega,s,\bm{k}_{\perp}}(u), \label{spatial plane wave sol} \\
    \psi^{+}_{\Omega,s,\bm{k}_{\perp}}(u) &= H^{+}_{\Omega} W_{s} + \frac{1}{\kappa}(\slashed{k} + \ell) \gamma^{0} H^{-}_{\Omega} W_{s}, \label{spin sol} \\
    \slashed{k} &= k_{a}\gamma^{a} =-(k^{1}\gamma^{1}+k^{2}\gamma^{2}), \notag 
\end{align}
where $\mathcal{N}_{\Omega,\bm{k}_{\perp}}$ is the spin-independent normalization constant  \cite{suzuki_analytic_2003,ueda_entanglement_2021},
\begin{align}
    W_{+} = 
    \begin{pmatrix*}[r]
    1 \\
    0 \\
    -1 \\
    0
    \end{pmatrix*}, \quad 
    W_{-} = 
    \begin{pmatrix*}[r]
    0 \\
    1 \\
    0 \\
    1
    \end{pmatrix*}, \label{general sol basis}
\end{align}
and
\begin{align}
    H^{\pm}_{\Omega}(u) &= H^{(1)}_{i\Omega \pm 1/2}(i\kappa u), \label{dirac hankel}
\end{align}
are the Hankel functions of the first kind $H^{(1)}_{\nu}(z)$ \cite{NIST:DLMF}. With the Rindler space Dirac inner product \eqref{dirac inner product}, the normalization constant is (see Appendix \ref{appendix: normalization rindler wavefunction})
\begin{align}
    \mathcal{N}_{\Omega,\bm{k}_{\perp}} = \mathcal{N}_{\Omega,+,\bm{k}_{\perp}} = \mathcal{N}_{\Omega,-,\bm{k}_{\perp}} &= \frac{1}{2\pi} \left( \frac{\kappa \cosh{(\pi \Omega)}}{8 e^{\pi \Omega}} \right)^{1/2},
\end{align}
and eigenspinor \eqref{full spin sol} obeys
\begin{align}
    i\hbar\partial_{v} \Psi^{+}_{\Omega,s,\bm{k}_{\perp}} &= \hbar \Omega \Psi^{+}_{\Omega,s,\bm{k}_{\perp}}.
\end{align}
Note that due to the form of the order $\nu$ and argument $z$ of our Hankel function \eqref{dirac hankel}, we have
\begin{align}
    (H^{\pm}_{\Omega})^{\ast} &= -H^{\mp}_{\Omega}. \label{hankel identity}
\end{align}
Also, if we let $\bm{k}_{\perp}= (0,0,0)$ and $\mathcal{N}_{\Omega,\bm{k}_{\perp}} \rightarrow 2\pi \mathcal{N}_{\Omega,\bm{k}_{\perp}=0}$ in eigenspinor \eqref{full spin sol}, the result is the solution for the linear Rindler Hamiltonian \eqref{rind ham op form} \cite{ueda_entanglement_2021}. It should be noted that the modified Bessel functions of the second kind $K_{i\Omega \pm 1/2}(\kappa u)$ (with its respective normalization constant) can also be used as solutions (see Refs.~\cite{suzuki_analytic_2003,rohim_relativistic_2021,ueda_entanglement_2021,crispino_unruh_2008}).

To find the energy levels using boundary condition \eqref{boulanger bc}, we use eigenspinor \eqref{spin sol} to identify $F$ and $G$ in the spinor decomposition \eqref{boulanger spinor ansatz} which are \cite{boulanger_bound_2006}
\begin{align}
    F \propto H^{-}_{\Omega} + s H^{+}_{\Omega}, \quad G \propto H^{-}_{\Omega} - s H^{+}_{\Omega}.
\end{align}
Then we use boundary condition \eqref{boulanger bc} along with identity \eqref{hankel identity} to get the spin-dependent quantization condition
\begin{align}
    \text{Re}\left[H^{(1)}_{i\Omega + 1/2}(i\mu)\right] + s \text{Im}\left[H^{(1)}_{i\Omega + 1/2}(i\mu)\right] = 0, \label{energy condition}
\end{align}
where $\mu=\kappa u_{0}$ is fixed. Note that the quantization condition \eqref{energy condition} can be derived using the MIT boundary conditions with chiral angles $\theta_{\text{M}}=0$ and $\theta_{\text{M}}=\pi$ for $s=+$ and $s=-$, respectively \cite{rohim_relativistic_2021}. Since $\mu \gg 1$, we see that the $\Omega$-zeros of the Hankel function for a large, fixed argument $\mu$ will satisfy the boundary condition. For the case $k=0$ and $s=+$, the zeros are given in \cite{rohim_relativistic_2021} which uses the numerical approximation scheme in \cite{FERREIRA2008223} to get an asymptotic expansion in $\mu_{0}=\ell u_{0}$  
\begin{align}
    \Omega_{n} &= \frac{\omega_{n} c}{g} \approx \mu_{0} - \frac{1}{2}+ a_{n+1} 2^{-1/3} \mu_{0}^{1/3} + \frac{a^{2}_{n+1}}{60}2^{1/3} \mu_{0}^{-1/3}  \notag \\
    &+ \frac{a_{n+1}}{6}2^{-1/3} \mu_{0}^{-2/3}  + \left( \frac{1}{70} - \frac{a^{3}_{n+1}}{700} - \frac{1}{12} \right) \mu_{0}^{-1}  \notag \\
    &+ O(\mu_{0}^{-4/3}), \label{hankel frequency zeros}
\end{align}
or
\begin{align}
    E_{n} &= \frac{\hbar g\Omega_{n}}{c} \approx mc^{2} + mgx_{0}a_{n+1} - \frac{\hbar g}{2c} +\frac{mg^{2}x^{2}_{0}}{30 c^{2}} a^{2}_{n+1} \notag \\
    &+ \frac{\hbar g^{2} x_{0}}{6 c^{3}}a_{n+1} + \frac{2mg^{3}x^{3}_{0}}{c^{4}} \left( \frac{1}{70} - \frac{a^{3}_{n+1}}{700} - \frac{1}{12} \right) \notag \\
    &+ O(c^{-5}), \label{hankel energy zeros}
\end{align}
for small $a_{n+1}$ where $a_{n+1}$ are the $(n+1)$-th zeros of the Airy function for $n=0,1,2,\ldots$. For a neutron, we have $\mu_{0} \sim 10^{31} \gg 1$ so eigenenergies \eqref{hankel frequency zeros} are an accurate approximation for the zeros of the Hankel function \cite{cochran_zeros_1965,FERREIRA2008223}. 

Following Ref.~\cite{rohim_relativistic_2021}, we use the kinetic energy $E^{D}_{n}=E_{n} -mc^{2}$ of Eq.~\eqref{hankel energy zeros} up to $O(c^{-3})$ to find the physically measureable energy level difference $E^{D}_{n^{\prime},n}$ between the $n$-th and $n^{\prime}$-th eigenstates 
\begin{align}
    E^{D}_{n^{\prime},n} &= E^{D}_{n^{\prime}} - E^{D}_{n}  \approx mgx_{0} (a_{n^{\prime}+1} - a_{n+1}) \notag \\
    &+ \frac{mg^{2}x^{2}_{0}}{30 c^{2}} \left(a^{2}_{n^{\prime}+1} - a^{2}_{n+1}\right) + \frac{\hbar g^{2}x_{0}}{6c^{3}} (a_{n^{\prime}+1}-a_{n+1}),
\end{align}
then the transition frequency is
\begin{align}
    \omega^{D}_{n^{\prime},n} &=\frac{E^{D}_{n^{\prime},n}}{\hbar} \approx \frac{mgx_{0}}{\hbar} (a_{n^{\prime}+1} - a_{n+1}) \notag \\
    &+ \frac{mg^{2}x^{2}_{0}}{30 \hbar c^{2}} \left(a^{2}_{n^{\prime}+1} - a^{2}_{n+1}\right) + \frac{g^{2}x_{0}}{6c^{3}} (a_{n^{\prime}+1}-a_{n+1}). \label{dirac trans freq}
\end{align}
To see the relativistic contributions to the nonrelativistic bouncing ball energies
\begin{align}
    E^{NR}_{n} &= mgx_{0}a_{n+1},
\end{align}
we use the nonrelativistic transition frequency
\begin{align*}
    \omega^{NR}_{n+1,n} &= \frac{E^{NR}_{n+1,n}}{\hbar}= \frac{mgx_{0}}{\hbar}\left(a_{n+2} - a_{n+1}\right),
\end{align*}
to define the transition frequency difference $\Delta\omega_{n+1,n}$ between $\omega^{D}_{n+1,n}$ and $\omega^{NR}_{n+1,n}$ \cite{rohim_relativistic_2021}
\begin{align}
    &\Delta\omega_{n+1,n} =\omega^{D}_{n+1,n} - \omega^{NR}_{n+1,n} \notag \\
    &\approx \frac{mg^{2}x^{2}_{0}}{30 \hbar c^{2}} \left(a^{2}_{n+2} - a^{2}_{n+1}\right) + \frac{g^{2}x_{0}}{6c^{3}} (a_{n+2}-a_{n+1}).
\end{align}
Now the neutron mass and gravitational acceleration on Earth's surface are, respectively, $m_{n} \approx 0.94$ GeV$/c^{2}$ and $g\approx 9.81$ $\text{m}/\text{s}^{2}$ so we find that 
\begin{align}
    \omega^{D}_{n+1,n}, \, \Delta\omega_{n+1,n} &\sim 10^{-20} \, \text{Hz}, \label{rel correction transition freq}
\end{align}
which is far too small to be detected using current technology given the sensitivity level of $\Delta \omega \sim 10^{-1}$ Hz measured in the \qb experiment \cite{cronenberg_acoustic_2018}. 

In Eq.~\eqref{hankel energy zeros}, the third term is interpreted to be the energy contribution from spin-gravity coupling
\begin{align}
    \frac{1}{c}\mathbf{S} \cdot \mathbf{g}, \label{spin gravity coupling term}
\end{align}
which does not appear in our FW Hamiltonian \eqref{nonrelativistic 3d rind ham}. While we demonstrated in Sec.~\ref{section: relativistic dfeg master equation} that our Dirac DFEG model does not affect spin, it is worth noting that previous literature has proven inconclusive as to the physical nature and relevance of the spin-gravity term. Initially, Peres \cite{peres_test_1978} proposed a simple ad-hoc model that modified the Dirac Lagrangian to include a spin-gravity coupling term with a dimensionless coupling constant $k$. Obukhov \cite{obukhov_spin_2001} later identified that $k=1/2$ by using an ``exact'' FW transformation that reproduced the spin-gravity coupling term \eqref{spin gravity coupling term}. However, subsequent work by Silenko and Teryaev \cite{silenko_semiclassical_2005} demonstrated that one could choose unitary transformations that could remove the term \eqref{spin gravity coupling term} through repeated FW transformations. This mathematical technicality has brought into question whether the FW transformation accurately provides physically relevant results since different unitary operators yield different results. Recent experiments on the equivalence principle using different spin orientations of spin-$1/2$ fermions \cite{duan_test_2016} yielded null results for spin-gravity coupling while another experiment \cite{tarallo_test_2014} provided an upper limit of $10^{-7}$ Hz for spin-$1/2$ fermions. Since the coupling term \eqref{spin gravity coupling term} is $g/(2c) \sim 10^{-8}$ Hz, experiments do not yet definitively prove nor disprove the existence of spin-gravity coupling. Coupled with the mathematical and physical ambiguity of the FW transformation, the question of spin-gravity coupling remains open, but we note that the appearance of the spin-gravity coupling energy in the energy levels \eqref{hankel energy zeros} of the full Rindler Hamiltonian \eqref{3d dirac ham op} (and equivalently the full observer's Hamiltonian \eqref{3d non FW ham}) lends some theoretical credence to its existence. Our work avoids this ambiguity with the spin-gravity term since this term only arises from the full Rindler Hamiltonian \eqref{3d dirac ham op} while our work uses the low energy gravitational Hamiltonian \eqref{low energy rind ham}.

\section{Normalization of the Rindler Wave Function} \label{appendix: normalization rindler wavefunction}

In this section, we calculate the normalization constant $\mathcal{N}_{\Omega,\bm{k}_{\perp}}$ of the eigenspinors \eqref{spin sol}. We will focus on the spin-up $s=+$ constant $\mathcal{N}_{\Omega,+,\bm{k}_{\perp}}$ then show that the constant is spin-independent, i.e., $\mathcal{N}_{\Omega,\bm{k}_{\perp}} = \mathcal{N}_{\Omega,+,\bm{k}_{\perp}} = \mathcal{N}_{\Omega,-,\bm{k}_{\perp}}$. We suppress the superscript $+$ and later on, we will also suppress the subscripts $s$ and $\bm{k}_{\perp}$ such that $\psi_{\Omega} \equiv \psi^{+}_{\Omega,+,\bm{k}_{\perp}}(u)$ and $\mathcal{N}_{\Omega,s,\bm{k}_{\perp}} \equiv \mathcal{N}_{\Omega}$. 

%In this section, we calculate the normalization constant $\mathcal{N}_{\Omega,\bm{k}_{\perp}}$ of the eigenspinors \eqref{spin sol}. The normalization of $\bm{k}_{\perp}$ is trivial so we will focus on the spin-up constant $\mathcal{N}_{\Omega,+,\bm{k}_{\perp}}$ of the spatial eigenspinor $\psi^{+}_{\Omega,+,\bm{k}_{\perp}}(u)$ then show that the constant is spin-independent, i.e., $\mathcal{N}_{\Omega,+,\bm{k}_{\perp}} = \mathcal{N}_{\Omega,-,\bm{k}_{\perp}}$. We suppress the superscript $+$ and subscripts $s$ and $\bm{k}_{\perp}$ such that $\psi_{\Omega} \equiv \psi^{+}_{\Omega,+,\bm{k}_{\perp}}(u)$ and $\mathcal{N}_{\Omega,s,\bm{k}_{\perp}} \equiv \mathcal{N}_{\Omega}$. 

If we use the spatial eigenspinor \eqref{spatial plane wave sol} with the Dirac inner product \eqref{dirac inner product}, we get
\begin{align}
    &\braket{f_{\Omega,s,\bm{k}_{\perp}} | f_{\Omega^{\prime},s^{\prime},\bm{k}^{\prime}_{\perp}}} = \int^{\infty}_{0} du \int^{\infty}_{-\infty} d^{2}x \, f^{\dagger}_{\Omega,s,\bm{k}_{\perp}} f_{\Omega^{\prime},s^{\prime},\bm{k}^{\prime}_{\perp}} \notag \\
    % &= \mathcal{N}_{\Omega,\bm{k}_{\perp}} \mathcal{N}_{\Omega^{\prime},\bm{k}^{\prime}_{\perp}} \int^{\infty}_{0} du \int^{\infty}_{-\infty} d^{2}x \,  \psi^{\dagger}_{\Omega,s,\bm{k}_{\perp}} \,\psi_{\Omega^{\prime},s^{\prime},\bm{k}^{\prime}_{\perp}} e^{i(k^{\prime} - k)x} \notag \\
    &= \overline{\mathcal{N}}_{\Omega,\bm{k}_{\perp}} \overline{\mathcal{N}}_{\Omega^{\prime},\bm{k}_{\perp}} \delta(\bm{k}_{\perp} - \bm{k}^{\prime}_{\perp}) \int^{\infty}_{0} du \, \psi^{\dagger}_{\Omega,s,\bm{k}_{\perp}} \,\psi_{\Omega^{\prime},s^{\prime},\bm{k}_{\perp}} \notag \\
    &= \overline{\mathcal{N}}_{\Omega,\bm{k}_{\perp}} \overline{\mathcal{N}}_{\Omega^{\prime},\bm{k}_{\perp}} \delta(\bm{k}_{\perp} - \bm{k}^{\prime}_{\perp}) \delta_{s,s^{\prime}} \braket{\psi_{\Omega,s,\bm{k}_{\perp}} | \psi_{\Omega^{\prime},s,\bm{k}_{\perp}}}, \\
    &d^{2}x = dx^{1} \, dx^{2}, \notag \\
    &\overline{\mathcal{N}}_{\Omega,\bm{k}_{\perp}} = 2\pi \mathcal{N}_{\Omega,\bm{k}_{\perp}}, \notag
\end{align}
where we used
\begin{align}
    &\int^{\infty}_{-\infty} d^{2}x \, e^{i(k^{\prime} - k)x} \equiv \int^{\infty}_{-\infty} d^{2}x \, e^{i(\bm{k}^{\prime}_{\perp} - \bm{k}_{\perp}) \cdot \mathbf{x}} \notag \\
    &= (2 \pi)^{2} \delta(\bm{k}_{\perp} - \bm{k}^{\prime}_{\perp}), \\
    &\overline{\mathcal{N}}_{\Omega^{\prime},\bm{k}^{\prime}_{\perp}} \psi^{\dagger}_{\Omega^{\prime},s^{\prime},\bm{k}^{\prime}_{\perp}}  \delta(\bm{k}_{\perp} - \bm{k}^{\prime}_{\perp}) \delta_{s,s^{\prime}} \notag \\
    &= \overline{\mathcal{N}}_{\Omega^{\prime},\bm{k}_{\perp}} \psi^{\dagger}_{\Omega^{\prime},s,\bm{k}_{\perp}} \delta(\bm{k}_{\perp} - \bm{k}^{\prime}_{\perp}) \delta_{s,s^{\prime}}.
\end{align}
Thus we only have to compute 
\begin{align}
    \braket{\psi_{\Omega,s,\bm{k}_{\perp}} | \psi_{\Omega^{\prime},s,\bm{k}_{\perp}}} &= \int^{\infty}_{0} du \, \psi^{\dagger}_{\Omega,s,\bm{k}_{\perp}} \,\psi_{\Omega^{\prime},s,\bm{k}_{\perp}}. \label{linear inner product}
\end{align}
As mentioned earlier, we will compute the spin-up $s=+$ constant and suppress the subscripts $s$ and $\bm{k}_{\perp}$ such that $\psi_{\Omega} \equiv \psi_{\Omega,s,\bm{k}_{\perp}}(u)$ and $\overline{\mathcal{N}}_{\Omega} \equiv \overline{\mathcal{N}}_{\Omega,\bm{k}_{\perp}}$. Additionally, note that $\psi_{\Omega}$ now satisfies the eigenvalue equation
\begin{align}
    &\hat{H}_{R,u} \psi_{\Omega} = \hbar \Omega \psi_{\Omega}, \label{op eig} \\
    &\hat{H}_{R,u} = \hbar u \left(\alpha_{1} k^{1} + \alpha_{2} k^{2} - i\alpha_{3} \partial_{u} \right) -\frac{i\hbar}{2}\alpha_{3} + \beta mcu. \label{plane wave rind ham}
\end{align}

We first derive the Lagrange-Green identity \cite{haberman_applied_2013} which will be crucial in calculating the normalization constant. For any arbitrary four-component spinors $\Psi_{1} \equiv \Psi_{1}(u)$ and $\Psi_{2} \equiv \Psi_{2}(u)$, consider the expression
\begin{equation}
    (\hat{H}_{R,u}\Psi_{1})^{\dagger}\Psi_{2}, \label{first expression}
\end{equation}
We expand the expression \eqref{first expression} using Eq.~\eqref{plane wave rind ham} to get
\begin{align}
    &(\hat{H}_{R,u}\Psi_{1})^{\dagger}\Psi_{2} = \left( \hbar u \left(\alpha_{1} k^{1} + \alpha_{2} k^{2} \right) \Psi_{1} \right)^{\dagger} \Psi_{2} \notag \\
    &+  \left(-i\hbar u\alpha_{3} \partial_{u} \Psi_{1}\right)^{\dagger}\Psi_{2} + \left(- \frac{i\hbar}{2}\alpha_{3}\Psi_{1} \right)^{\dagger}\Psi_{2} \notag \\
    &+ (\beta mcu \Psi_{1})^{\dagger}\Psi_{2} \notag \\
    &=  \left(-i\hbar u\alpha_{3} \partial_{u} \Psi_{1}\right)^{\dagger}\Psi_{2} + \Psi^{\dagger}_{1}\left[ \vphantom{\frac{1}{1}} \hbar u \left(\alpha_{1} k^{1} + \alpha_{2} k^{2} \right) \right.\notag \\
    &\left. + \frac{i\hbar}{2}\alpha_{3} + \beta mcu \right] \Psi_{2}. \label{op expand}
\end{align}
If we use the identity
\begin{align}
    \partial_{u} (i\hbar \Psi_{1}^{\dagger} \alpha_{3} u \Psi_{2}) &= (\partial_{u} \Psi_{1}^{\dagger} ) (i\hbar\alpha_{3}u\Psi_{2}) \notag \\
    &+ \Psi_{1}^{\dagger}\left(i\hbar\alpha_{3} \partial_{u} (u\Psi_{2})\right),
\end{align}
the first term in the second equality of Eq.~\eqref{op expand} is
\begin{align}
    &\left(-i\hbar u\alpha_{3} \partial_{u} \Psi_{1}\right)^{\dagger}\Psi_{2} =  \partial_{u} (i\hbar \Psi_{1}^{\dagger} \alpha_{3} u \Psi_{2}) \notag \\
    &- \Psi_{1}^{\dagger}\left(i\hbar\alpha_{3} \partial_{u} (u\Psi_{2})\right), \notag \\
    &=\partial_{u} (i\hbar \Psi_{1}^{\dagger} \alpha_{3} u \Psi_{2}) + \Psi_{1}^{\dagger}\left[ -i\hbar\alpha_{3} u \partial_{u} - i\hbar \alpha_{3} \right] \Psi_{2}.
\end{align}
Combining our results then yields the differential Lagrange-Green identity
\begin{align}
    (\hat{H}_{R,u} \Psi_{1})^{\dagger}\Psi_{2} = \partial_{u} (i\hbar \Psi_{1}^{\dagger} \alpha_{3} u \Psi_{2}) + \Psi_{1}^{\dagger}(\hat{H}_{R,u}\Psi_{2}), \label{lg ident}
\end{align}
which leads to the integral Lagrange-Green identity \cite{haberman_applied_2013}
\begin{align}
     \int^{\infty}_{0} du \, (\hat{H}_{R,u} \Psi_{1})^{\dagger}\Psi_{2} &= i\hbar \Psi_{1}^{\dagger} \alpha_{3} u \Psi_{2} \Big\lvert^{\infty}_{0} \notag \\
     &+ \int^{\infty}_{0} du \, \Psi_{1}^{\dagger}(\hat{H}_{R,u}\Psi_{2}), \label{int lg ident}
\end{align}
when integrated according to the inner product \eqref{linear inner product}. 

Now let $\Psi_{1}=\psi_{\Omega}$ and $\Psi_{2}=\psi_{\Omega^{\prime}}$ then we have
\begin{align}
    \int^{\infty}_{0} du \, (\hat{H}_{R,u} \psi_{\Omega})^{\dagger} \psi_{\Omega^{\prime}}
     &= i\hbar \psi^{\dagger}_{\Omega} \alpha_{3} u \psi_{\Omega^{\prime}} \Big\lvert^{\infty}_{0} \notag \\
     &+ \int^{\infty}_{0} du \, \psi^{\dagger}_{\Omega}(\hat{H}_{R,u}\psi_{\Omega^{\prime}}). \label{lagrange green with psi}
\end{align}
Since $\psi_{\Omega}$ is an eigenfunction of $\hat{H}_{R,u}$, it obeys the eigenvalue equation \eqref{op eig} so Eq.~\eqref{lagrange green with psi} becomes
\begin{align}
    \hbar\Omega\int^{\infty}_{0} du \, \psi^{ \dagger}_{\Omega}\psi_{\Omega^{\prime}}
     &= i\hbar \psi^{ \dagger}_{\Omega} \alpha_{3} u \psi_{\Omega^{\prime}} \Big\lvert^{\infty}_{0} \notag \\
     &+ \hbar \Omega^{\prime}\int^{\infty}_{0} du \, \psi^{ \dagger}_{\Omega}\psi_{\Omega^{\prime}},
\end{align}
which yields after rearrangement \cite{soffel_dirac_1980,greiner_quantum_1985}
\begin{align}
    \Braket{\psi_{\Omega} | \psi_{\Omega^{\prime}}} &= \int^{\infty}_{0} du \, \psi^{ \dagger}_{\Omega}\psi_{\Omega^{\prime}} \notag \\
    &= \frac{iu}{\Omega - \Omega^{\prime}}  \, \psi^{ \dagger}_{\Omega} \alpha_{3} \psi_{\Omega^{\prime}} \Big\lvert^{\infty}_{0}. \label{pre norm1}
\end{align}
To evaluate the right hand side of \eqref{pre norm1}, we first expand the expression using the Hankel functions in spinor \eqref{spin sol} to get
\begin{align}
    &\frac{i u}{\Omega - \Omega^{\prime}} \, \psi^{ \dagger}_{\Omega} \alpha_{3} \psi^{ \dagger}_{\Omega^{\prime}} \Big\lvert^{\infty}_{0} \notag \\
    &= \frac{2iu}{\Omega - \Omega^{\prime}} \Big( (H^{-}_{\Omega})^{\ast} H^{-}_{\Omega^{\prime}} - (H^{+}_{\Omega})^{\ast} H^{+}_{\Omega^{\prime}} \Big) \Big\lvert^{\infty}_{0} \notag \\
    &= \frac{2i u}{\Omega - \Omega^{\prime}} \Big(H^{(2)}_{-i\Omega - 1/2}(-i\kappa u) H^{(1)}_{i\Omega^{\prime} - 1/2}(i\kappa u) \notag \\
    &- H^{(2)}_{-i\Omega + 1/2}(-i\kappa u) H^{(1)}_{i\Omega^{\prime} + 1/2}(i\kappa u) \Big) \Big\lvert^{\infty}_{0}, \label{pre norm2}
\end{align}
where instead of using identity \eqref{hankel identity}, we have elected to use \cite{NIST:DLMF}
\begin{align}
    \Big(H^{(1)}_{\nu}(z) \Big)^{\ast} &= H^{(2)}_{\nu^{\ast}}(z^{\ast}),
\end{align}
with $H^{(2)}_{\nu}(z)$ being the Hankel functions of the second kind. For large arguments $z$, the Hankel functions of the first and second kind have the asymptotic expansion \cite{NIST:DLMF}
\begin{align}
    H^{(1)}_{\nu}(z) &\sim \sqrt{\frac{2}{\pi z}} e^{+ i\left(z - \frac{1}{2}\nu\pi - \frac{1}{4}\pi \right)}, \\
    H^{(2)}_{\nu}(z) &\sim \sqrt{\frac{2}{\pi z}} e^{- i\left(z - \frac{1}{2}\nu\pi - \frac{1}{4}\pi \right)},
\end{align}
respectively, for $z,\nu \in \mathbb{C}$, which vanishes as $z \rightarrow \infty$ so the right hand side of Eq.~\eqref{pre norm1} vanishes as $u \rightarrow \infty$. Thus the only nontrivial limit to consider is $u \rightarrow 0$. Due to the Hankel function's singularity at $u=0$, we utilize the modified Bessel functions of the second kind $K_{\nu}(z)$ which is related to the Hankel functions by \cite{NIST:DLMF}
\begin{align}
    K_{\nu}(z) = \begin{cases}
        \phantom{+}\frac{i\pi}{2} e^{+i\nu \pi/2}H^{(1)}_{\nu}(ze^{+i\pi/2}), &-\pi \leq \text{ph}z \leq \frac{\pi}{2}, \\
        -\frac{i\pi}{2} e^{-i\nu \pi/2}H^{(2)}_{\nu}(ze^{-i\pi/2}), &-\frac{\pi}{2} \leq \text{ph}z \leq \pi,
    \end{cases} \label{mb2 lim}
\end{align}
Then as $z \rightarrow 0$, the modified Bessel functions of the second kind $K_{\nu}(z)$ have the asymptotic form \cite{NIST:DLMF}
\begin{align}
    K_{\nu}(z) \sim \frac{\Gamma(\nu)}{2} \left({\frac{z}{2}}\right)^{-\nu} = \frac{2^{\nu-1}}{z^{\nu}} \Gamma(\nu), \label{mb2 to han}
\end{align}
where $\Gamma(z)$ is the gamma function. We will first evaluate the second term in Eq.~\eqref{pre norm2}. Using Eqs.~\eqref{mb2 lim} and \eqref{mb2 to han} in the limit $u \rightarrow 0$, we have
\begin{align}
    &\frac{2i u}{\Omega - \Omega^{\prime}} H^{(2)}_{-i\Omega + 1/2}(-i\kappa u) H^{(1)}_{i\Omega^{\prime} + 1/2}(i\kappa u) \notag \\
    &\xrightarrow[u\rightarrow 0]{} \left(\frac{4i}{\kappa\pi^{2} \Delta\Omega} \right) \left(\frac{2}{\kappa} \right)^{-i\Delta \Omega} \left(\frac{1}{u} \right)^{-i\Delta \Omega} e^{\pi(\Omega + \Omega^{\prime})/2} \notag \\ 
    &\Gamma(-i\Omega + 1/2) \Gamma(i\Omega^{\prime} + 1/2) \notag \\
    &=\left(\frac{4i}{\kappa\pi^{2}} \right) \left(\frac{2}{\kappa} \right)^{-i\Delta \Omega} e^{\pi(\Omega + \Omega^{\prime})/2} \Gamma(-i\Omega + 1/2) \notag \\
    &\Gamma(i\Omega^{\prime} + 1/2) \left( \frac{\cos{[(\Delta \Omega) x]}-i\sin{[(\Delta \Omega) x]}}{\Delta \Omega} \right), \label{pre norm3}
\end{align}
where $\Delta \Omega= \Omega-\Omega^{\prime}$, $x=\ln{(1/u)}$, and we have used the identities
\begin{align}
    \left(\frac{1}{u}\right) ^{-i\Delta \Omega} &= e^{-i(\Delta \Omega)\ln(1/u)} = e^{-i(\Delta \Omega) x}, \\
    e^{iz} &= \cos{(z)} + i \sin{(z)}.
\end{align}
Since $x=\ln{(1/u)}$ diverges rapidly as $u \rightarrow \infty$, we can use the following identities
\begin{align}
    \lim_{x \to \infty} \frac{\sin{[(\Delta \Omega) x]}}{\Delta \Omega} &= \pi \delta(\Omega- \Omega^{\prime}), \\
    \lim_{x \to \infty} \frac{\cos{[(\Delta \Omega) x]}}{\Delta \Omega} &= 0,
\end{align}
in Eq.~\eqref{pre norm3} to finally get 
\begin{align}
    &\lim_{u \to 0} \left(\frac{2iu}{\Omega - \Omega^{\prime}} H^{(2)}_{-i\Omega + 1/2}(-i\kappa u) H^{(1)}_{i\Omega^{\prime} + 1/2}(i\kappa u) \right) \notag \\
    &= -\left(\frac{4}{\kappa\pi} \right) \left(\frac{2}{\kappa} \right)^{-i\Delta \Omega} e^{\pi(\Omega + \Omega^{\prime})/2} \Gamma(-i\Omega + 1/2) \notag \\
    &\Gamma(i\Omega^{\prime} + 1/2) \delta(\Omega - \Omega^{\prime}) \notag \\
    &= -\frac{4 e^{\pi \Omega}}{\kappa\pi} \abs{\Gamma(i\Omega + 1/2)}^{2} \delta(\Omega - \Omega^{\prime}) \notag \\
    &=-\frac{4 e^{\pi \Omega}}{\kappa\cosh{(\pi\Omega)}} \delta(\Omega - \Omega^{\prime}),
\end{align}
where we used the identities
\begin{align}
    \abs{\Gamma(1/2+it)}^{2} &= \frac{\pi}{\cosh{(\pi t)}}, \\
    f(y)\delta(y-t) &= f(t), \quad y,t \in \mathbb{R}, \label{delta function identity}
\end{align}
in the last equality. Repeating the same procedure for the first term in Eq.~\eqref{pre norm2} yields
\begin{align}
    &\lim_{u \to 0} \left(\frac{2i u}{\Omega - \Omega^{\prime}} H^{(2)}_{-i\Omega - 1/2}(-i\kappa u) H^{(1)}_{i\Omega^{\prime} - 1/2}(i\kappa u) \right) \notag \\
    &=\frac{4 e^{\pi \Omega}}{\kappa\cosh{(\pi\Omega)}}  \delta(\Omega - \Omega^{\prime}).
\end{align}
Then the delta-normalized inner product is
\begin{align}
    \Braket{\psi_{\Omega} | \psi_{\Omega^{\prime}}} &= \int^{\infty}_{0} du \, \psi^{ \dagger}_{\Omega} \psi_{\Omega^{\prime}}\notag \\
    &=\frac{8 e^{\pi \Omega}}{\kappa\cosh{(\pi\Omega)}}  \delta(\Omega - \Omega^{\prime}),
\end{align}
so
\begin{align}
    \abs{\overline{\mathcal{N}}_{\Omega}}^{2} &= (2\pi)^{2} \abs{\mathcal{N}_{\Omega}}^{2} = \frac{\kappa \cosh{(\pi \Omega)}}{8 e^{\pi \Omega}},
\end{align}
thus the spin-up $s=+$ normalization constant is
\begin{align}
    \mathcal{N}_{\Omega} &= \frac{1}{2\pi} \left( \frac{\kappa \cosh{(\pi \Omega)}}{8 e^{\pi \Omega}} \right)^{1/2}.
\end{align}
Repeating the same procedure for the spin-down $s=-$ spinor yields the same constant $\mathcal{N}_{\Omega,-,\bm{k}_{\perp}} = \mathcal{N}_{\Omega,+,\bm{k}_{\perp}}$ thus the spin-independent normalization constant is
\begin{align}
    \mathcal{N}_{\Omega,\bm{k}_{\perp}} = \mathcal{N}_{\Omega,+,\bm{k}_{\perp}} = \mathcal{N}_{\Omega,-,\bm{k}_{\perp}} = \frac{1}{2\pi} \left(\frac{\kappa \cosh{(\pi \Omega)}}{8 e^{\pi \Omega}} \right)^{1/2}. \label{hankel norm const}
\end{align}

To calculate the normalization constant $\mathcal{N}^{K}_{\Omega,\bm{k}_{\perp}}$ if one had used the modified Bessel functions of the second kind $K_{i\Omega \pm 1/2}(\kappa u)$ in solution \eqref{spin sol}, we use identity \eqref{mb2 lim} in solution \eqref{spin sol} and absorb the introduced constant terms into the normalization constant \eqref{hankel norm const} to get
\begin{align}
    \mathcal{N}^{K}_{\Omega,\bm{k}_{\perp}} =\mathcal{N}^{K}_{\Omega,+,\bm{k}_{\perp}} = \mathcal{N}^{K}_{\Omega,-,\bm{k}_{\perp}} = \frac{1}{2\pi} \left(\frac{\kappa \cosh{(\pi \Omega)}}{2 \pi^{2}} \right)^{1/2}.
\end{align}

\bibliography{gravityspinbib}

\end{document}